\documentclass[]{elsart}
\usepackage{epsfig}
\usepackage{amssymb}

\begin{document}
\begin{frontmatter}

\title{\bf Using the Scaling Analysis to Characterize Financial Markets}

\author[label1,label2]{T. Di Matteo},
\author[label2]{T. Aste},
\author[label3,cor1]{M. M. Dacorogna}
\address[label1]{INFM - Dipartimento di Fisica "E. R. Caianiello",
Universit\`a degli Studi di Salerno, 84081 Baronissi (SA),
Italy.}
\address[label2]{Applied Mathematics, Research School of Physical Sciences, Australian National University, 0200 Canberra, Australia.}
\corauth[cor1]{Corresponding author:Tel: +41 1 6399760, fax: +41 1 6399961.}
\ead{michel.dacorogna@converium.com}
\address[label3]{Converium Ltd, General Guisan - Quai 26, 8022 Zurich, Switzerland.}

\begin{abstract}
We empirically analyze the scaling properties of daily Foreign Exchange rates, Stock Market indices and Bond futures across different financial markets.
We study the scaling behaviour of the time series by using a generalized Hurst exponent approach.
We verify the robustness of this approach and we compare the results with the scaling properties in the frequency-domain.
We find evidence of deviations from the pure Brownian motion behavior. We show that these deviations are associated with characteristics of the specific markets and they can be, therefore, used to distinguish the different degrees of development of the markets.
\end{abstract}

\begin{keyword}
Scaling exponents; Econophysics; Time series analysis.
\\
\noindent {\it JEL Classification:} C00; C1; G00; G1.

\end{keyword}
\end{frontmatter}

\newpage

\section{Introduction}
\label{s.Introduction}

The scaling concept is increasingly applied outside the
traditional physical sciences domain
\cite{Muller90,BouchaudPot,Rebonato,Wilmott,Hull,LibrMant,LibDac}.
In the recent years, its application to financial markets,
initiated by Mandelbrot in the 1960~\cite{Mandelbrot,MandelbrotB},
has largely increased also in consequence of the abundance of
available data \cite{Muller90}. Two types of scaling behaviors are
studied in the finance literature:
\begin{enumerate}
\item The behavior of some forms of volatility measure (variance
of returns, absolute value of returns) as a function of the
\emph{time interval} on which the returns are measured. (This
study will lead to the estimation of a \emph{scaling exponent}
related to the Hurst exponent.)
\item The behavior of the tails of
the distribution of returns as a function of \emph{the size of the
movement} but keeping the time interval of the returns constant.
(This will lead to the estimation of the \emph{tail index} of the
distribution \cite{LibDac}.)
\end{enumerate}
Although related, these two analysis lead to different quantities
and should not be confused as it is often the case in the
literature as can be seen in the papers and debate published in
the November 2001 issue of Quantitative Finance. For more
explanations about this and the relation between the two
quantities, the reader is referred to the excellent paper by
Groenendijk {\it et al.}~\cite{Groenendijk}. In this study, we are
interested in the first type of analysis. Until now, most of the
work has concentrated in studies of particular markets: Foreign
Exchange \cite{Muller90,LibDac,Corsi01}, Stock \cite{MantSta95} or
Fixed Income \cite{Ballocchi}. These studies showed that empirical
scaling laws hold in all these markets and for a large range of
frequencies: from few minutes to few months. In a recent book
\cite{LibDac}, the hypothesis of heterogeneous market agents was
developed and backed by empirical evidences. In this view, the
agents are essentially distinguished by the frequency at which
they operate in the market. The scaling analysis, which looks at
the volatility of returns measured at different time intervals, is
a parsimonious way of assessing the relative impact of these
heterogeneous agents on price movements. Viewing the market
efficiency as the result of the interaction of ¨these agents
\cite{Daco01}, brings naturally to think that it is the presence
of many different agents that would characterize a mature market,
while the absence of some type of agents should be a feature of
less developed markets. Such a fact should then reflect in the
measured scaling exponents. The study of the scaling behaviors
must therefore be an ideal candidate to characterize markets. To
further explore this issue, we perform an empirical analysis of
daily data across different financial markets to examine the
similarities or differences in the scaling properties.

Until the 1960s, the only stochastic and scaling model in finance
was the Brownian motion, originally proposed by Bachelier in 1900
\cite{LibrMant,Bachelier1900}, and developed several decades later
\cite{Osborne59}. This theory predicts that the returns of market
prices should follow a normal distribution with stable mean and
finite variance. However, there are ample empirical evidences that
the returns are not normally distributed but have higher peak
around the mean and fatter tails \cite{LibrMant,LibDac}. Moreover,
it is also observed that volatility clustering is a general
characteristic of financial markets \cite{LibDac}. Generalizations
of the classical Brownian motion were made by Mandelbrot and
followers involving either fractional Brownian motions
\cite{MandelbrotB,Mand65,Mand68}, or L\'evy motion
\cite{Mandelbrot,Mand62,Fama63,Fama65,Mand67,Mirowski95}. Closely
related additive scaling models have also been developed
\cite{LibrMant,MantSta95,Evertsz95}: Brownian, fractional
Brownian, L\'evy processes. The above approaches generally involve
additive monofractal processes and analyses; but, in contrast,
several scaling systems appear to be more complex. Recently, a
controversy has erupted
\cite{BlakeLeBaron01,Lux01,Mandelbrot01,StanPlerou} to know if the
processes that describe financial data are truly scaling or simply
an artifact of the data. Moreover, several publications propose
new scaling models or empirical analyses that better describe
empirical evidences
\cite{BlakeLeBaron01,Lux01,Mandelbrot01,StanPlerou,Bouchaud2000}.
It should be however noted that - as underlined by Stanley et al.
\cite{Stanle96} - in statistical physics, when a large number of
microscopic elements interact without characteristic scale,
universal macroscopic scaling laws may be obtained independently
of the microscopic details.

In this paper we address the question of the scaling properties of
financial time series by empirically analyzing daily data for
Foreign Exchange rates, Stock Market indices and Bond futures
(described in Section~\ref{s.data}). We study very developed as
well as emerging markets in order to see if the scaling properties
differ between the two and if they can serve to characterize and
measure the development of the market. Here the scaling law is not
used to conclude anything on the theoretical process but on the
contrary we use it as a ``stylized fact'' that any theoretical
model should also reproduce. The purpose of this paper is to point
out how a relatively simple statistics gives us indications on the
market characteristics, very much along the lines of the review
paper by Brock \cite{ScalEco}. In Section~\ref{s.Scalinganalysis},
we recall the theoretical framework and we introduce the
generalized Hurst exponents analysis. In
Section~\ref{s.metodology} we describe the methodology utilized to
empirically analyze the data. In Section~\ref{ResDisc} we compute
and compare the scaling spectral exponents and the Hurst
exponents. Finally some conclusions are given in
Section~\ref{s.conclusion}.

\section{The Studied Markets}
\label{s.data}

We study several financial markets which are at different
development stage: mature and liquid markets, emerging and less
liquid markets. These markets deal with different instruments:
equities, foreign exchange rates, fixed income futures. In
particular, the data that we analyze are: Foreign Exchange rates
(FX) (see Table~\ref{t.1}), Stock Market indices (SM) (see
Table~\ref{t.2}), Treasury rates corresponding to twelve different
maturity dates (TR) (see Table~\ref{t.3}) and Eurodollar rates
having maturity dates ranging from $3$ months to $4$ years (ER)
(see Table~\ref{t.4}). Hereafter we give a brief description of
the time-series studied in this paper.

\begin{itemize}
\item {\bf FX:} The Foreign Exchange rates (Table~\ref{t.1}) are
daily rates of exchange of major currencies against the U.S.
dollar. The time series that we study go from $1990$ to $2001$ and
$1993$ to $2001$. These rates have been certified by the Federal
Reserve Bank of New York for customs purposes. The data are noon
buying rates in New York for cable transfers payable in the listed
currencies. These rates are also those required by the Securities
and Exchange Commission (SEC) for the integrated disclosure system
for foreign private issuers. The information is based on data
collected by the Federal Reserve Bank of New York from a sample of
market participants.

\item {\bf SM:} The Stock Market indices (reported in
Table~\ref{t.2}) are 32 of the major indices of both very
developed markets and emerging markets. These daily time series
range from $1990$ or $1993$ to $2001$.

\item {\bf TR:} The Treasury rates (Table~\ref{t.3}) are daily
time series going from $1990$ to $2001$. The yields on Treasury
securities at `constant maturity' are interpolated by the U.S.
Treasury from the daily yield curve. This curve, which relates the
yield on a security to its time to maturity, is based on the
closing market bid yields on actively traded Treasury securities
in the over-the-counter market. These market yields are calculated
from composites of quotations obtained by the FD Bank of New York.
The constant maturity yield values are read from the yield curve
at fixed maturities, currently 3 and 6 months and 1, 2, 3, 5, 7,
10, and 30 years. The Treasury bill rates are based on quotes at
the official close of the U. S. Government securities market for
each business day. They have maturities of 3 and 6 months and 1
year.

\item {\bf ER:} The Eurodollar interbank interest rates
(Table~\ref{t.4}) are bid rates with different maturity dates and
they are daily data in the time period
$1990$-$1996$~\cite{DiMatteo}.
\end{itemize}

\begin{figure}
\begin{center}
\mbox{\epsfig{file=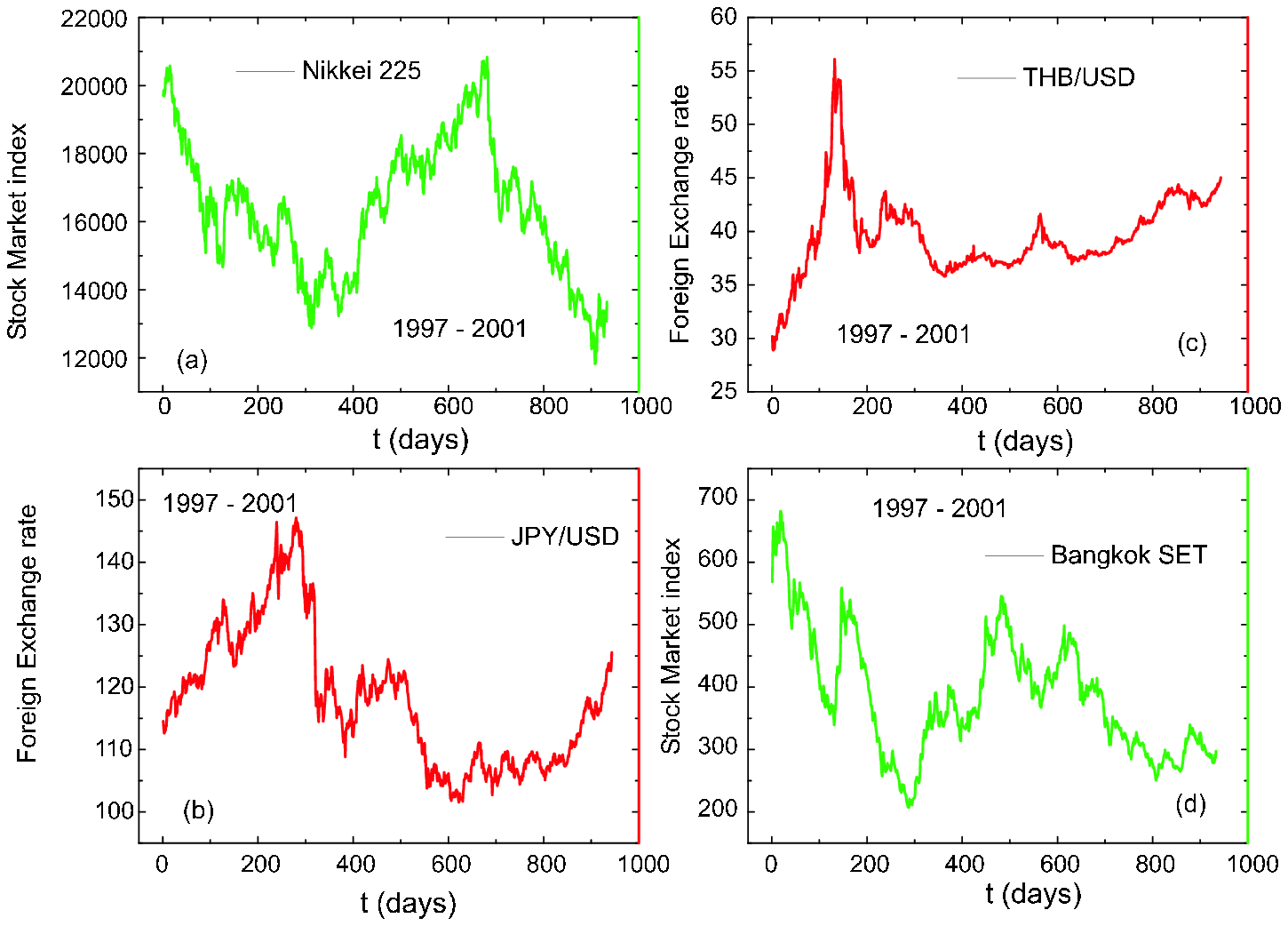,width=6.in,angle=0}}
\end{center}
\caption{The Foreign Exchange rates and the Stock Market indices as a function of time $t$ in the time period $1997$-$2001$; (a) Nikkei 225; (b) JPY/USD; (c) THB/USD; (d) Bangkok SET.}
\label{f.fig1}
\end{figure}

\begin{figure}
\begin{center}
\mbox{\epsfig{file=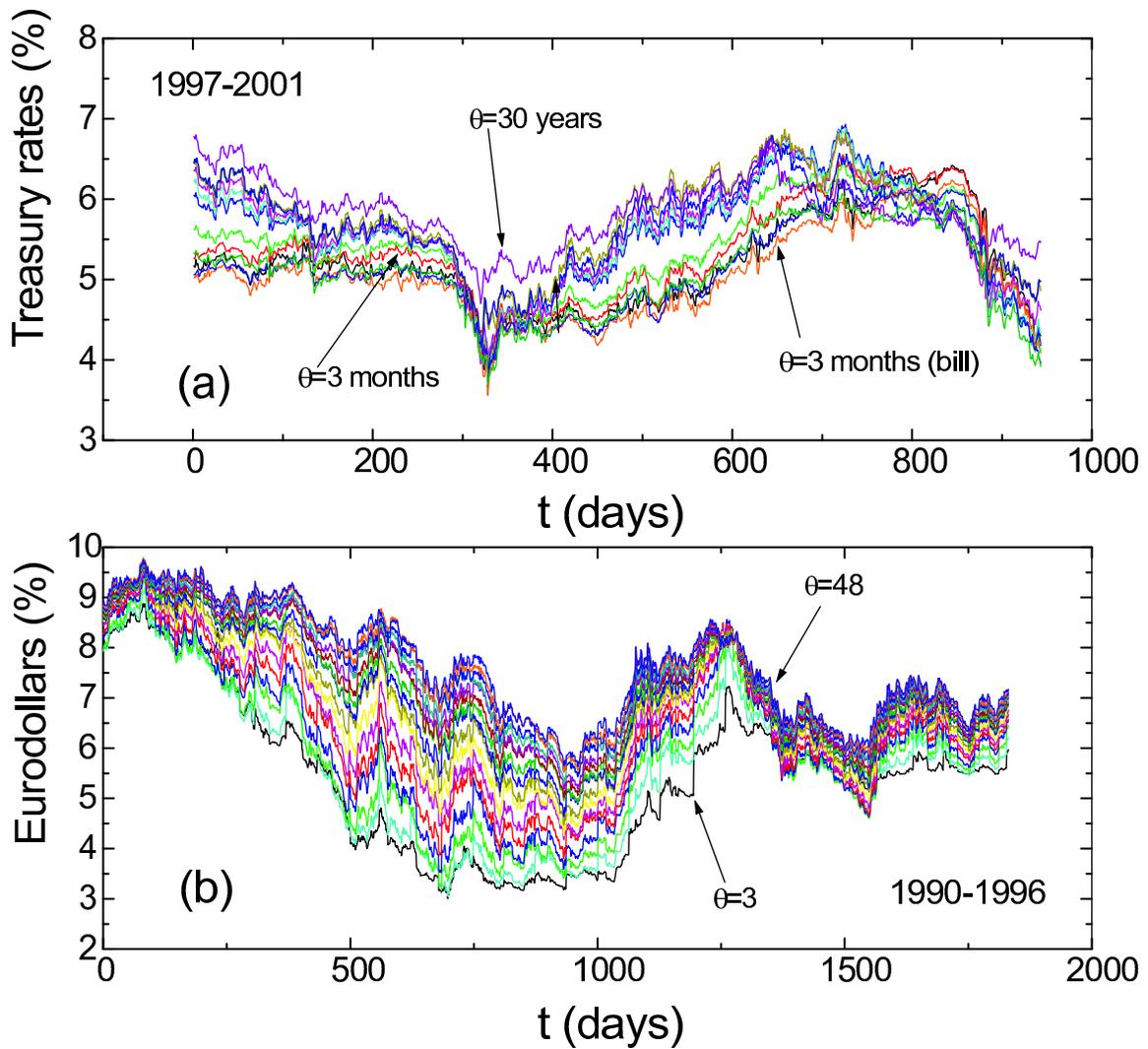,width=6.in,angle=0}}
\end{center}
\caption{(a) The Treasury rates at `constant maturity' as a function of $t$ in the time period $1997$-$2001$. Each curve corresponds to a maturity date $\theta$, ranging from $3$ months to $30$ years and Treasury bill rates to a maturity date $\theta$=$3$,$6$ months and $1$ year; (b) The Eurodollar interest rates as a function of $t$ in the time period $1990$-$1996$. Each curve corresponds to a maturity date $\theta$, ranging from $3$ months to $48$ months.}
\label{f.fig2}
\end{figure}

As an example, the behaviors of FX rates and the SM rates for
Japan (JPY/USD and Nikkei 225) and Thailand (THB/USD and Bangkok
SET) as a function of time $t$ are shown in Fig.~\ref{f.fig1} in
the time period $1997$-$2001$. Another example is given in
Fig.~\ref{f.fig2} that shows the TR time series as a function of
$t$ at different maturities dates in the time period $1997$-$2001$
and the ER time series as a function of $t$ at different
maturities dates in the time period $1990$-$1996$.

\section{The theoretical framework}
\label{s.Scalinganalysis}

The scaling properties in time series have been studied in the
literature by means of several techniques such as the rescaled
range statistical analysis $R/S$~\cite{HurstBlack65} and modified
$R/S$ analysis \cite{Lo91}, multiaffine analysis, detrended
fluctuation analysis (DFA)
\cite{Peng94,Stanley96,Viswa97,Hu01,Vandewalle97,Ausloos2000},
periodogram regression (GPH method) \cite{Geweke83}, the ARFIMA
estimation by exact maximum likelihood \cite{Sowell92,Carles2000},
the $(m,k)$-Zipf method \cite{Zipf}, the moving-average analysis
technique \cite{Ellinger71}, the Average Wavelet Coefficient
Method \cite{Mehrabi97,Simonsen98,Percival2000,Gencay2001}, and
-in this respect- connection to multi-fractal/multi-affine
analysis (the $q$ order height-height correlation) have been made
\cite{Vandewalle98,VandAus98,Ivanova99}. The rescaled range
statistical analysis ($R/S$ analysis) was first introduced by
Harold Edwin Hurst \cite{HurstBlack65} to describe the long-term
dependence of water levels in rivers and reservoirs. It provides a
sensitive method for revealing long-run correlations in random
processes. This analysis can distinguish random time series from
correlated time series and gives a measure of a signal
``roughness''. What mainly makes the Hurst analysis appealing is
that all these information about a complex signal are contained in
one parameter only: the \emph{Hurst exponent}. The original
approach of Hurst to the scaling properties of the time series is
recalled in details in Appendix. One of the weaknesses
of the original method is that it relies on maximum and minimum
data, which makes it very sensitive to outliers. In order to study
the multi-fractal features of the data we here use an alternative
method to the original approach of Hurst.

\subsection{Generalized Hurst exponent}
\label{s.Hurstgeneralized}

The Hurst analysis brings to light that some statistical
properties of time series $X(t)$ (with $t$=$\nu$,
$2\nu$,~...,~$k\nu$,~...,~$T$) scale with the observation-period
($T$) and the time-resolution ($\nu$). Such a scaling is
characterized by an exponent $H$ which is commonly associated with
the long-term statistical dependence of the signal. A
generalization of the approach proposed by Hurst should therefore
be associated with the scaling behavior of statistically
significant properties of the signal. To this purpose we analyze
the $q$-order moments of the distribution of the increments
\cite{MandelbrotB,Barabasi91} which is a good characterization of
the statistical evolution of a stochastic variable $X(t)$. It is
given by the:
\begin{equation}
\label{multi}
K_q(\tau) =
\frac{ \left< |X(t+\tau) - X(t)|^q \right> }{\left< |X(t)|^q \right> }\;\;\;,
\end{equation}
where the time-interval $\tau$ can vary between $\nu$ and
$\tau_{max}$. (Note that, for $q=2$, the $K_q(\tau)$ is
proportional to the autocorrelation function $a(\tau)=\left<
X(t+\tau)X(t)\right>$.)

\begin{figure}
\begin{center}
\mbox{\epsfig{file=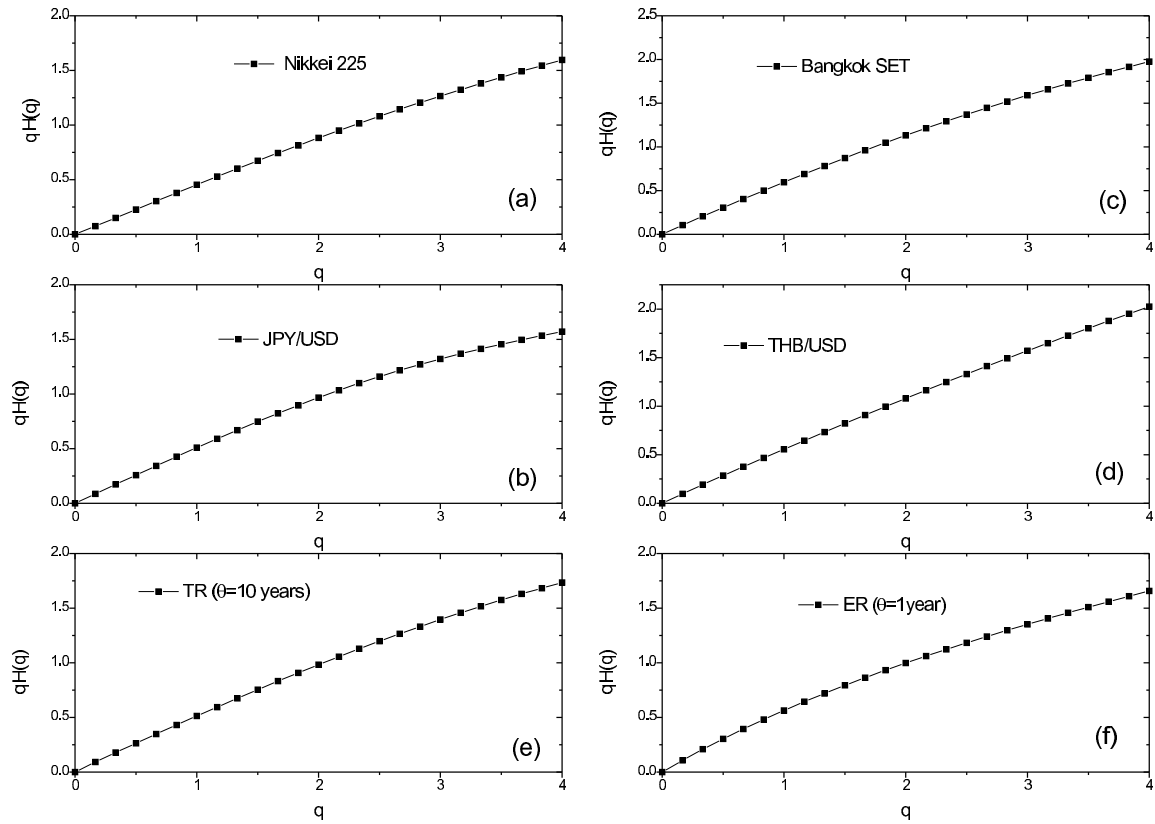,width=6.in,angle=0}}
\end{center}
\caption{The function $q H(q)$ vs. $q$ in the time period from $1997$ to $2001$. (a) JAPAN (Nikkei 225); (b) JAPAN (JPY/USD); (c) Thailand (Bangkok SET); (d) Thailand (THB/USD); (e) Treasury rates having maturity dates $\theta=10$ years; (f) Eurodollar rates having maturity dates $\theta=1$ year. For (f) the time period is $1990$ - $1996$.}
\label{f.HQT}
\end{figure}

The generalized Hurst exponent $H(q)$\footnote{We use $H$ without
parenthesis as the original Hurst exponent and $H(q)$ as the
generalized Hurst exponent.} can be defined from the scaling
behavior of $K_q(\tau)$ which can be assumed to be given by the
relation \cite{Barabasi91}
\begin{equation}
\label{Hurstgen}
K_q(\tau) \sim
\left( \frac{\tau }{ \nu } \right)^{qH(q)}\;\;\;.
\end{equation}
Within this framework, we can distinguish between two kind of
processes: (i) a process where $H(q) = H$, constant independent of
$q$; (ii) a process with $H(q)$ not constant. The first case is
characteristic of uni-scaling or uni-fractal processes and its
scaling behavior is determined from a unique constant $H$ that
coincides with the Hurst exponent $H$. This is for instance the
case for self-affine processes where $qH(q)$ is linear ($H(q)=H$)
and fully determined by its index $H$. (Recall that, a
transformation is called affine when it scales time and distance
by different factors, while a behavior that reproduces itself
under affine transformation is called self-affine
\cite{MandelbrotB}. A time-dependent self-affine function $X(t)$
has fluctuations on different time scales that can be rescaled so
that the original signal $X(t)$ is statistically equivalent to its
rescaled version $\lambda ^{-H} X(\lambda t)$ for any positive
$\lambda$ \cite{Feder88}, i.e. $X(t) \sim  \lambda ^{-H} X(\lambda
t)$. Brownian motion is self-affine by nature.) In the second
case, when $H(q)$ depends on $q$, the process is commonly called
multi-scaling (or multi-fractal) \cite{Feder88,West85} and
different exponents characterize the scaling of different
$q$-moments of the distribution. In this `curve' of exponents
$H(q)$, some values of $q$ are associated with special features.
For instance, when $q=1$, $H(1)$ describes the scaling behavior of
the absolute values of the increments. The value of this exponent
is expected to be closely related to the original Hurst exponent,
$H$, that is indeed associated with the scaling of the absolute
spread in the increments. The exponent at $q=2$, is associated
with the scaling of the autocorrelation function and is related to
the power spectrum \cite{Flandrin89}. A special case is associated
with the value of $q=q^*$ at which $q^*H(q^*)=1$. At this value of
$q$, the moment $K_{q^*}(\tau)$ scales linearly in $\tau$
\cite{MandelbrotB}. Since $qH(q)$ is \emph{in general} a monotonic
growing function of $q$, we have that all the moments $H_q(\tau)$
with $q < q^*$  will scale slower than $\tau$, whereas all the
moments with $q > q^*$ will scale faster than $\tau$. The point
$q^*$ is therefore a threshold value. In this paper we focalize
the attention on the case $q=1$ and $2$. Clearly in the
uni-fractal case $H(1)=H(2)=H(q^*)$. Their values will be equal to
1/2 for the Brownian motion and they would be equal to $H
\not=0.5$ for the fractional Brownian motion. However, for more
complex signals, these coefficients do not in general coincide. We
thus see that the non-linearity of the empirical function $qH(q)$
is a solid argument against Brownian, fractional Brownian, L\'evy,
and fractional L\'evy models, which are all additive models
therefore giving for $qH(q)$ straight lines or portions of
straight lines. The curves for $qH(q)$ vs. $q$ are reported in
Fig.~\ref{f.HQT} for some of the data. One can observe that, for
all these time series, $qH(q)$ is not linear in $q$ but slightly
bending below the linear trend. The same behavior holds for the
other data. This is a clear sign of deviation from Brownian,
fractional Brownian, L\'evy, and fractional L\'evy models, as
already seen in FX rates \cite{Muller90}. (Other cases showing
marked deviations from Brownian motion have been discussed
elsewhere
\cite{Vandewalle98,Ivanova99,1AuIva01,2AuIva01,VanAus98}.)

\subsection{Scaling spectral density and Hurst exponent}
\label{s.beta}

For financial time series, as well as for many other stochastic
processes, the spectral density $S(f)$ is empirically found to
scale with the frequency $f$ as a power law: $S(f) \propto
f^{-\beta}$. Here we use a simple argument to show how this
scaling in the frequency domain is related to the scaling in the
time-domain. Indeed, it is known that the spectrum $S(f)$ of the
signal $X(t)$ can be conveniently calculated from the Fourier
transform of the autocorrelation function (Wiener-Khinchin
theorem). On the other hand, the autocorrelation function of
$X(t)$ is proportional to the second moment of the distribution of
the increments which, from Eq.~\ref{Hurstgen}, is supposed to
scale as $K_2 \sim \tau^{2H(2)}$. But, the components of the
Fourier transform of a function which behaves in the time-domain
as $\tau^{\alpha}$ are proportional to $f^{-\alpha-1}$ in the
frequency-domain. Therefore, we have that the power spectrum of a
signal that scales as Eq.~\ref{Hurstgen} must behave as:
\begin{equation}
\label{www}
S(f) \propto f^{-2H(2)-1} \;\;\;.
\end{equation}
Consequently, the slope $\beta$ of the power spectrum is related
with the generalized Hurst exponent for $q=2$ through:
$\beta=1+2H(2)$ \cite{Flandrin89}. Note that Eq.~\ref{www} is
obtained only assuming that the signal $X(t)$ has a scaling
behavior in accordance to Eq.~\ref{Hurstgen}. Here we are not
making any hypothesis on the kind of underlying mechanism that
might lead to such a scaling behavior.

\section{Methodology}
\label{s.metodology}

Let us here recall that the theoretical framework we presented in
the previous section is based on the assumption that the signal
has the scaling property described in Eq.~\ref{Hurstgen}.
Moreover, we have implicitly assumed that the scaling properties
associated with a given time series stay unchanged across the
observation time window $T$. On the other hand, it is well known
that financial time series show evidences of variation of their
statistical properties with time, and depend on the observation
time window $T$. The simplest case which shows such a dependence
is the presence of a linear drift ($ \eta t $) added to a
stochastic signal ($X(t)= \tilde X(t)+\eta t$) with $\tilde X(t)$
satisfying Eq.~\ref{Hurstgen} and the above mentioned properties
of stability within the time window. Clearly, the scaling analysis
described in the previous section must be applied to the
stochastic component $\tilde X(t)$ of the signal. This means that
we must subtract the drift $\eta t$ from the signal $X(t)$. To
this end one can evaluate $\eta$ from the following relation:
\begin{equation}
\label{avinBeta}
\left< X(t+\tau) - X(t) \right> = \eta \tau \;\;\;.
\end{equation}
Other more complex deviations from the stationary behavior might
be present in the financial data that we analyze. In this context,
the subtraction of the linear drift can be viewed as a first
approximation. The accuracy of this approximation has been
verified by varying the observation time window. We verify that
the results obtained for different time window sizes are all
comparable, with fluctuations within a range of $10 \%$.

Our empirical analysis is performed on the daily time series TR,
ER, FX and SM (described in Section~\ref{s.data}) which span
typically over periods between $1000$ and $3000$ days. In
particular, we analyze the time series themselves for the TR and
ER, whereas we compute the returns from the logarithmic price
$X(t)=\ln(P(t))$ for FX and SM. Moreover, all of these variables
are `detrended' by eliminating the linear drift as described above
(Eq.~\ref{avinBeta}).

We compute the $q$-order moments $K_q(\tau)$ (defined in
Eq.~\ref{multi}) of the `detrended' signals and their logarithms
with $\tau$ in the range between $\nu=1$ day and $\tau_{max}$
days. In order to test the robustness of our empirical approach,
for each series we analyze the scaling properties varying
$\tau_{max}$ between $5$ and $19$ days. The resulting exponents
computed using these different $\tau_{max}$ are stable in their
values within a range of $10 \%$.

We verify that the scaling behavior given in Eq.~\ref{Hurstgen} is
well followed and we compute the associated generalized Hurst
exponent $H(q)$ whose values are given in Section~\ref{ResDisc}.

In order to test that our method is not biased we estimate the
generalized Hurst exponents for simulated random walks produced by
using three different random numbers generators. We perform $100$
simulations of random walks with the same number of data points as
in our samples ($991$ and $3118$) and estimate the generalized
Hurst exponents $H(1)$ and $H(2)$ and the power spectra exponents
$\beta$. The results are reported in Table~\ref{t.HbetaGaussian}.
In all the cases, $H(1)$ and $H(2)$ have values of $0.5$ within
the errors. Only when we consider uniformly distributed random
numbers in the interval ($0$,$1$) (Rand uses a lagged Fibonacci
generator combined with a shift register random integer generator,
based on the work of Marsaglia.) we obtain for $H(1)$ of $0.47 \pm
0.01$, but also in this case $H(2)$ is 0.5 within the errors. This
shows that our method is powerful and robust and does not suffer of
bias as other methods do. On the other hand, the estimations of
$\beta$ from the power spectrum have values around $1.8$ (instead
of $2$), showing therefore that this other method is affected by a
certain bias.

\section{Results and Discussion}
\label{ResDisc}

\subsection{Computation of the generalized Hurst exponent}

\begin{figure}
\begin{center}
\mbox{\epsfig{file=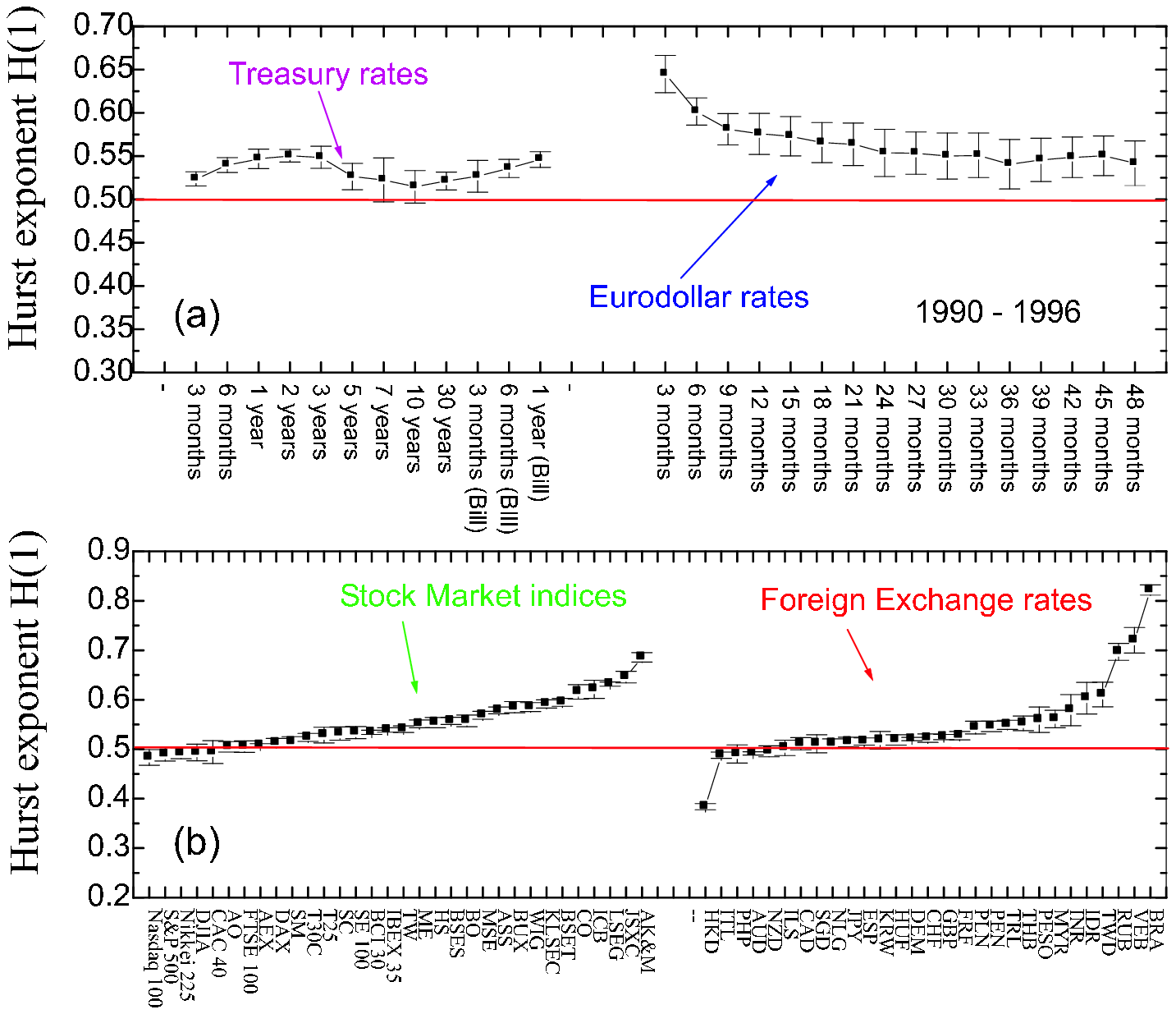,width=6.in,angle=0}}
\end{center}
\caption{(a) The Hurst exponent $H(1)$ for the Treasury and Eurodollar rates time series in the period from $1990$ to $1996$; (On the $x$-axis the corresponding maturities dates are reported.) (b) The Hurst exponent $H(1)$ for the Stock Market indices and Foreign Exchange rates in the time period reported in Tabs. ~\ref{t.1} and ~\ref{t.2}. (On the $x$-axis the corresponding data-sets are reported.)}
\label{f.H1log}
\end{figure}

\begin{figure}
\begin{center}
\mbox{\epsfig{file=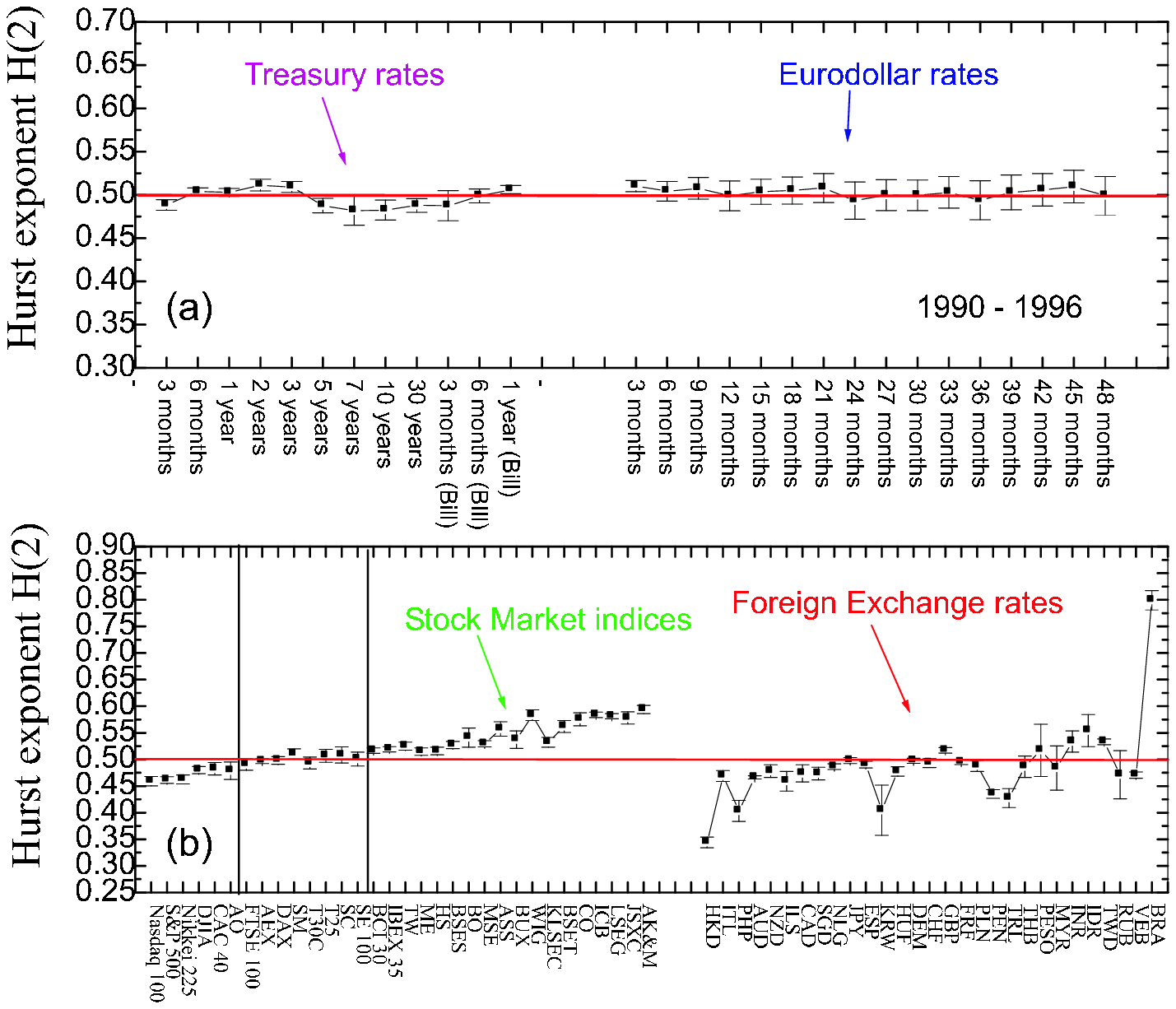,width=6.in,angle=0}}
\end{center}
\caption{(a) The Hurst exponent $H(2)$ for the Treasury and Eurodollar rates time series in the period from $1990$ to $1996$; (On the $x$-axis the corresponding maturities dates are reported.) (b) The Hurst exponent $H(2)$ for the Stock Market indices and Foreign Exchange rates in the time period reported in Tabs. ~\ref{t.1} and ~\ref{t.2}. (On the $x$-axis the corresponding data-sets are reported.)}
\label{f.H2log}
\end{figure}

In this section we report and discuss the results for the scaling
exponents $H(q)$ computed for $q=1$ and $q=2$. These exponents
$H(1)$ and $H(2)$ for all the assets and different markets
(presented in Section~\ref{s.data}) are reported in
Figs.~\ref{f.H1log}~and~\ref{f.H2log} respectively.
Figures~\ref{f.H1log}~(a)~and~\ref{f.H2log}~(a) refer to the
Treasury and Eurodollar rates in the time period from $1990$ to
$1996$. Whereas Figures~\ref{f.H1log}~(b)~and~\ref{f.H2log}~(b)
are relative to the Stock Market indices and Foreign Exchange
rates in the time period reported in
Tables~\ref{t.1}~and~\ref{t.2}. The data points are the average
values of $H(1)$ and $H(2)$ computed from a set of values
corresponding to different $\tau_{max}$ and the error bars are
their standard deviations. Let us first observe that, for fixed
income instruments
(Figs.~\ref{f.H1log}~(a)~and~\ref{f.H2log}~(a)), $H(2)$ is close
to $0.5$ while $H(1)$ is rather systematically above $0.5$ (with
the $3$ months Eurodollar rate that shows a more pronounced
deviation because it is directly influenced by the actions of
central banks). On the other hand, when Stock markets are
concerned, we find that the generalized Hurst exponents $H(1)$,
$H(2)$ show remarkable differences between developed and emerging
markets. In particular, the values of $H(1)$, plotted in
Fig.~\ref{f.H1log} (b), present a differentiation across $0.5$
with high values of $H(1)$ associated with the emerging markets
and low values of $H(1)$ associated with developed ones. Moreover,
we can see from Fig.~\ref{f.H2log} (b) that the different assets
can be classified into three different categories: First the ones
that have an exponent $H(2)>0.5$ which includes all indices of the
emerging markets and the BCI $30$ (Italy), IBEX $35$ (Spain) and
the Hang Seng (Hong Kong). A second category concerns the data
exhibiting $H(2) \sim  0.5$ (within the error bars). This category
includes: FTSE $100$ (UK), $AEX$ (Netherlands), $DAX$ (Germany),
Swiss Market (Switzerland), Top $30$ Capital (New Zealand),
Telaviv $25$ (Israel), Seoul Composite (South Korea) and Toronto
SE $100$ (Canada). A third category is associated with $H(2)<0.5$
and includes the following data: Nasdaq $100$ (US), $S\&P 500$
(US), Nikkei $225$ (Japan), Dow Jones Industrial Average (US), CAC
$40$ (France) and All Ordinaries (Australia). We find therefore
that all the emerging markets have $H(2) \geq  0.5$ whereas all
the well developed have $H(2) \leq 0.5$.

For what concerns the Foreign exchange rates, we find that they
show $H(1)>0.5$ quite systematically which is consistent with
previous results computed with high frequency data
\cite{Muller90}, although the values here are slightly lower. An
exception with pronounced $H(1)<0.5$ is the HKD/USD (Hong Kong)
(Fig. ~\ref{f.H1log}~(b)). This FX rate is or has been at one
point pegged to the USD, that is why its exponent differs from the
others. Whereas in the class $H(1) \sim  0.5$ we have: $ITL/USD$
(Italy), $PHP/USD$ (Philippines), $AUD/USD$ (Australia), NZD/USD
(New Zealand), ILS/USD (Israel), $CAD/USD$ (Canada), $SGD/USD$
(Singapore), $NLG/USD$ (Netherlands) and $JPY/USD$ (Japan). On the
other hand the values of $H(2)$ (Fig.~\ref{f.H2log}~(b)) show a
much larger tendency to be $<0.5$ with some stronger deviations
such as: $HKD/USD$ (Hong Kong), $PHP/USD$ (Philippines), $KRW/USD$
(South Korea), $PEN/USD$ (Peru) and $TRL/USD$ (Turkey). Whereas
values of $H(2)>0.5$ are found in: $GBP/USD$ (United Kingdom),
$PESO/USD$ (Mexico), $INR/USD$ (India), $IDR/USD$ (Indonesia),
$TWD/USD$ (Taiwan) and $BRA/USD$ (Brazil).

Let us remind that $H(2)>0.5$ is commonly associated with a
persistent behavior in the fluctuations of the returns, whereas an
exponent $H(2)<0.5$ indicates anti-persistence \cite{Feder88}.

\begin{figure}
\begin{center}

\mbox{\epsfig{file=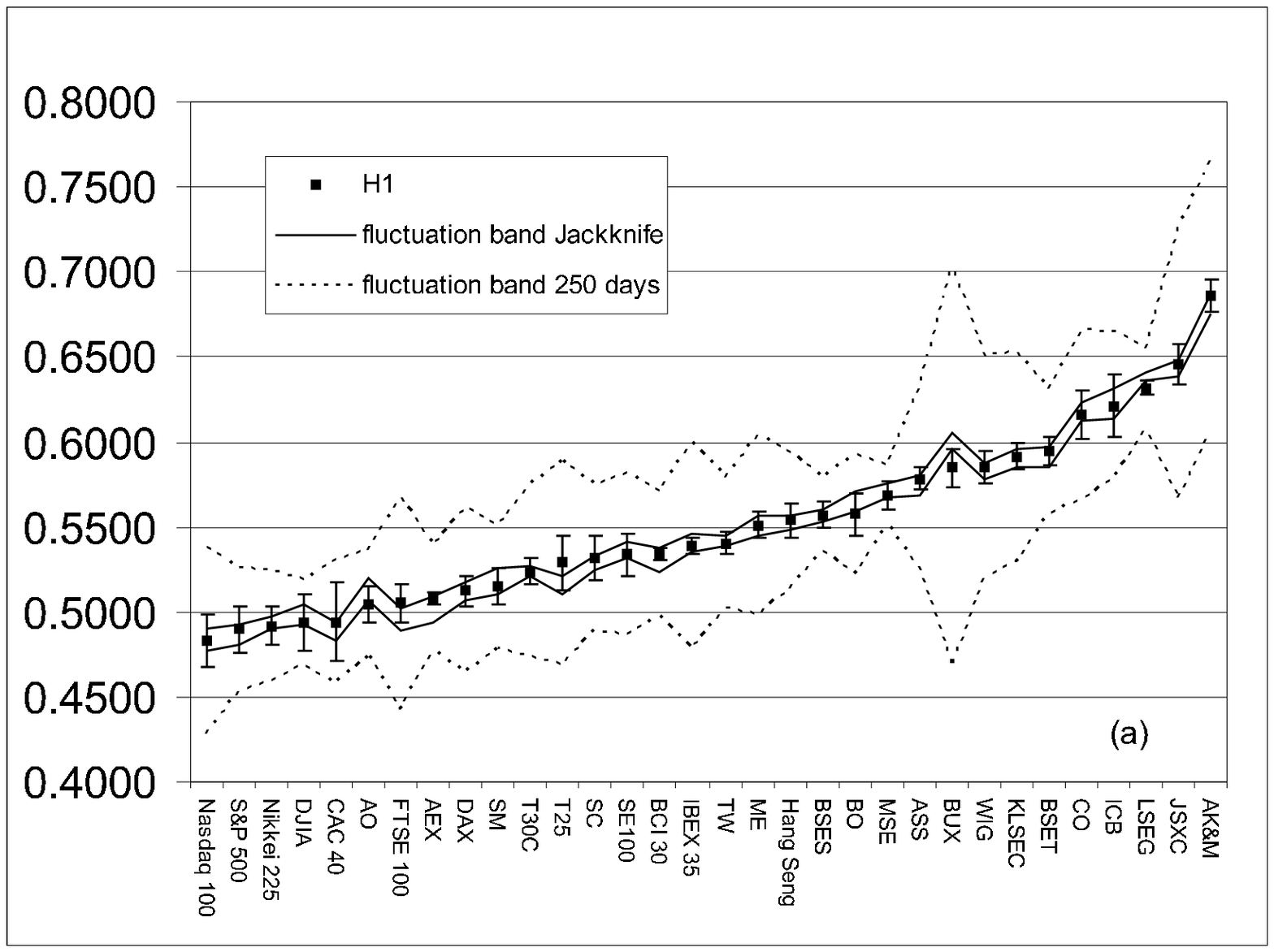,width=4.in,angle=0}}
\mbox{\epsfig{file=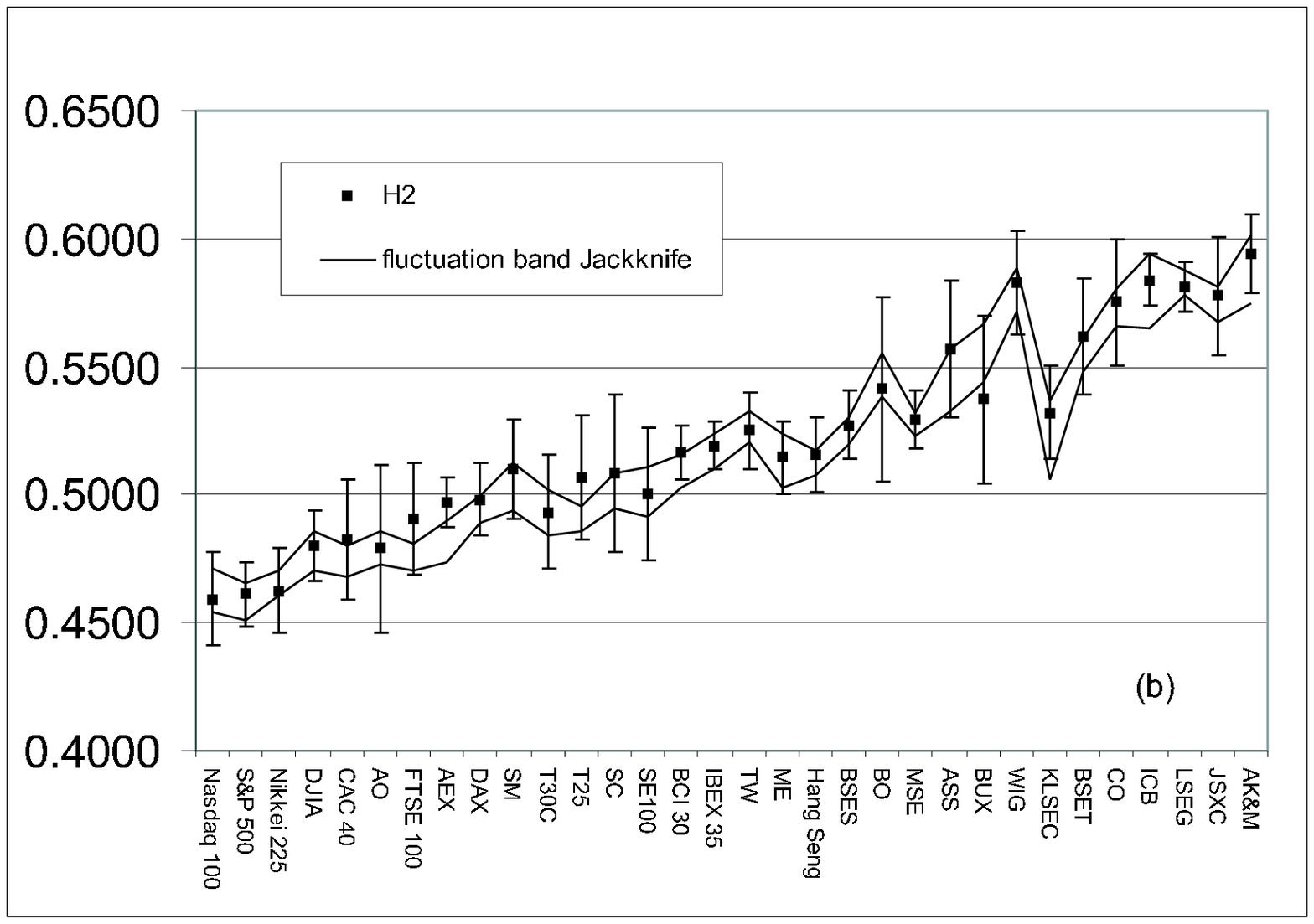,width=4.in,angle=0}}
\end{center}
\caption{(a) The generalized Hurst exponent $H(1)$ for the Stock Market indices in the whole time period (see Tab. ~\ref{t.2}) with its variation (black lines) obtained by using the Jackknife method and its variation (dashed lines) when time periods of $250$ days are considered; (b) The generalized Hurst exponent $H(2)$ for the Stock Market indices in the whole time period (see Tab. ~\ref{t.2}) with its variation (black lines) obtained by using the Jackknife method. The square points are the average values of $H(1)$ and $H(2)$ computed from a set of values corresponding to different $\tau_{max}$. The error bars are their standard deviations.}
\label{f.Fig6}
\end{figure}

These analysis has been done also on different time periods and
the values are reported in Table~\ref{t.Htimeperiod9701} for the
exponents $H(1)$ and $H(2)$ for the time period from $1997$ to
$2001$ for Foreign Exchange rates, Stock Market indices and
Treasury rates. In order to verify the stability of these results
over different time periods we calculate and compare, for the
Stock market data, the generalized Hurst exponents $H(1)$ and
$H(2)$ in the whole time period (shown in Table~\ref{t.2}) and in
time periods of $250$ days. Moreover, we tested the numerical
robustness of our results by using the Jackknife method
\cite{kunsch89} which consists in taking out randomly $1/10$ of
the sample and iterates the procedure $10$ times (every time
taking out data which were not taken out in the previous runs). We
observe (see Fig.~\ref{f.Fig6}) that the generalized Hurst
exponents computed on these Jackknife-reduced time series are very
close to those computed on the entire series with deviations
inside the errors estimated by varying $\tau_{max}$ (as described
in Section~\ref{s.metodology}). On the other hand, the analysis on
sub-periods of $250$ days shows fluctuations that are larger than
the previous estimated errors (and larger than the variations with
the Jackknife method) indicating therefore that there are
physically-significant changes in the market behaviors over
different time periods (Fig.~\ref{f.Fig6}~(a)). This phenomenon
was also detected in \cite{LibDac} when studying Exchange rates
that were part of the European Monetary System. It seems that
$H(1)$ is particularly sensitive to institutional changes on the
market. This study confirms it for Stock indices. The scaling
exponents cannot be assumed to be constant over time if a market
is experiencing major institutional changes. Nevertheless, well
developed markets have values of $H(2)$ that are on average
smaller than the emerging ones. Moreover, the weakest markets have
oscillation bands that stay above $0.5$ whereas the strongest have
oscillation bands that contain $0.5$.

\subsection{Spectral analysis}

\begin{figure}
\begin{center}
%\mbox{\epsfig{file=FigSpettroFX.eps,width=6.in,angle=0}}
\mbox{\epsfig{file=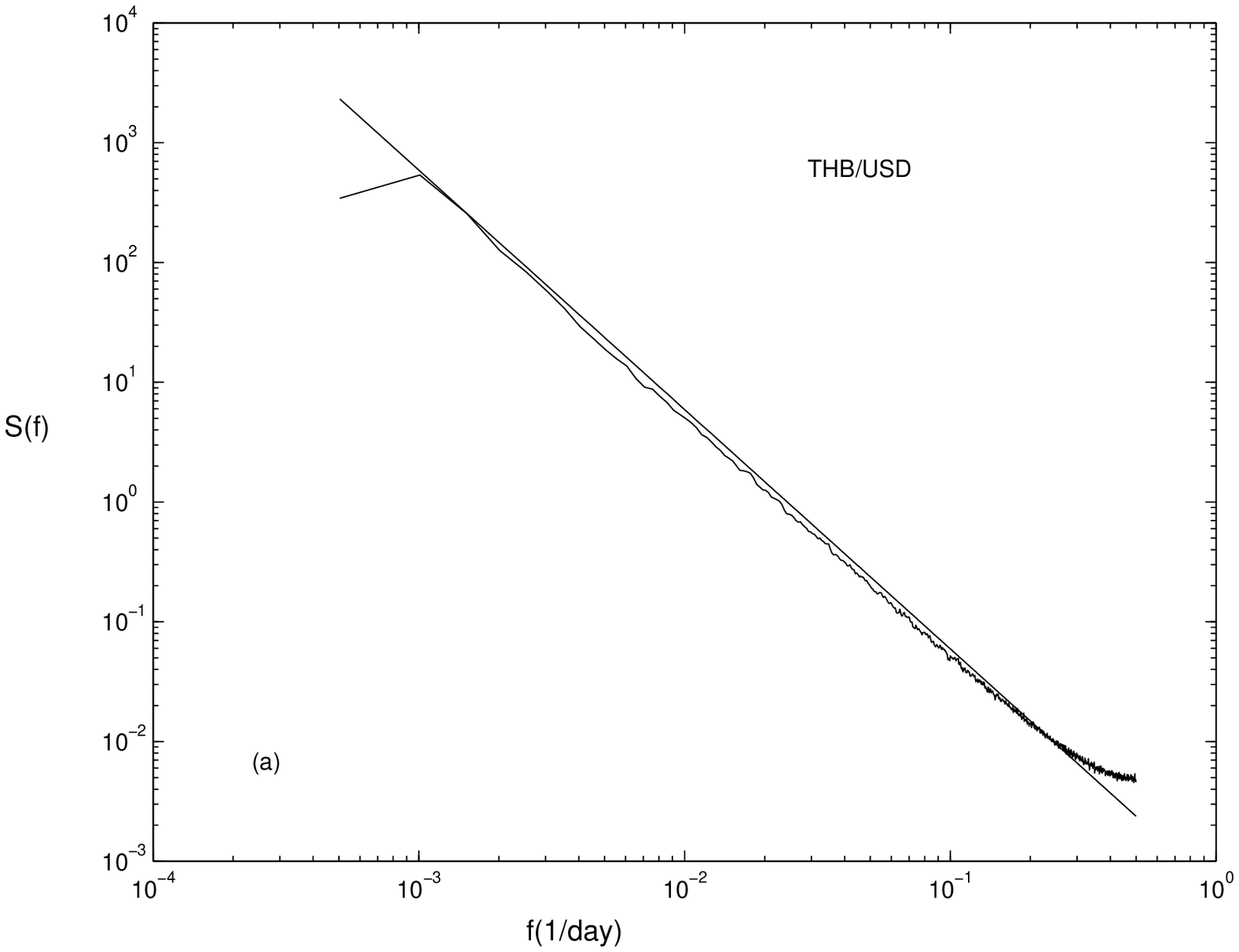,width=3.in,angle=0}}
\mbox{\epsfig{file=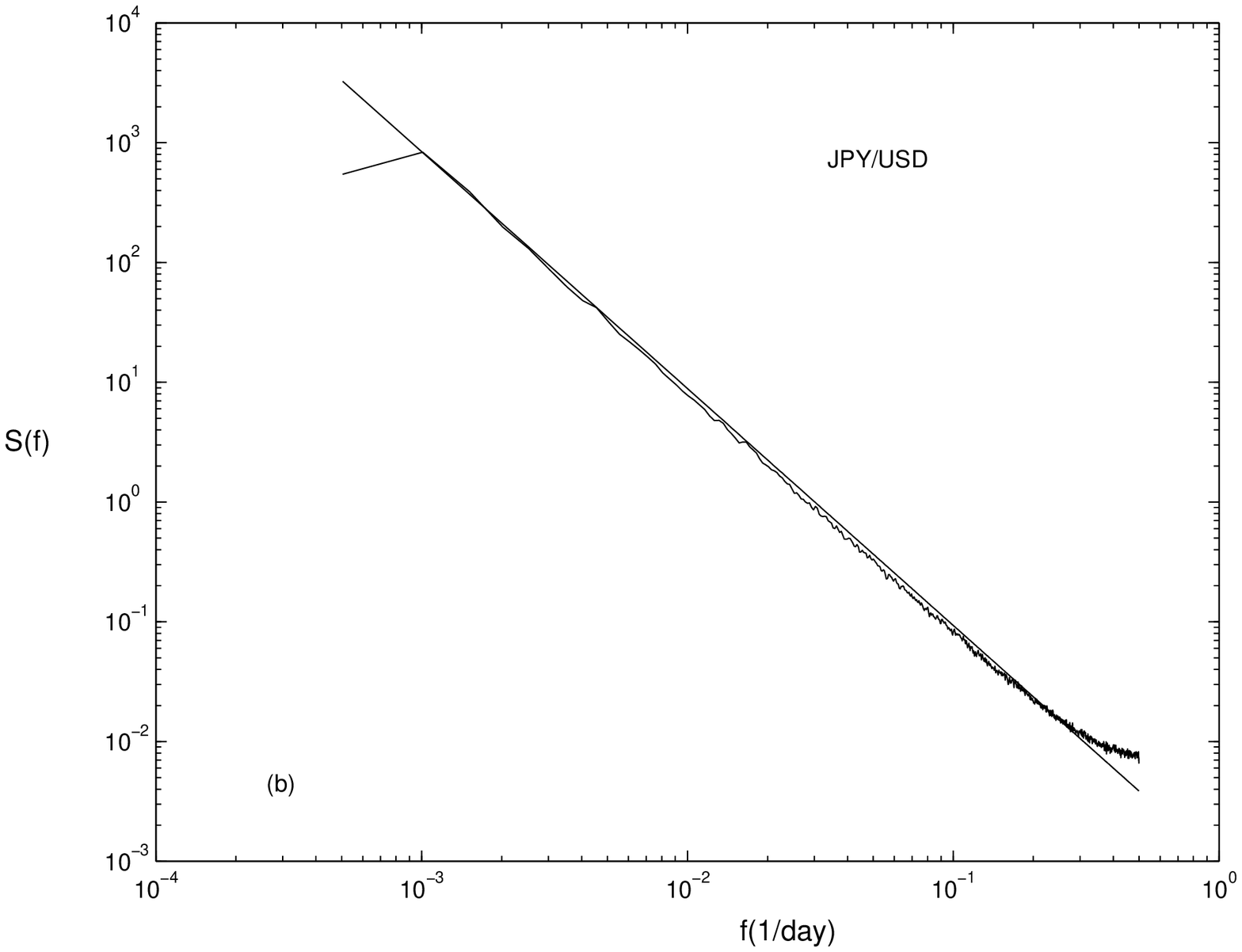,width=3.in,angle=0}}
\end{center}
\caption{The power spectra of the Foreign Exchange rates compared with the behaviour of $f^{-2H(2)-1}$ (straight lines in log-log scale) computed using the Hurst exponents values in the time period $1997$-$2001$; (a) Thailand (THB/USD) and (b) JAPAN (JPY/USD). The line is the prediction from the generalized Hurst exponent $H(2)$ (Eq. ~\ref{www}).}
\label{f.SpettroFX}
\end{figure}

\begin{figure}
\begin{center}
%\mbox{\epsfig{file=FigSpettroSM.eps,width=6.in,angle=0}}
\mbox{\epsfig{file=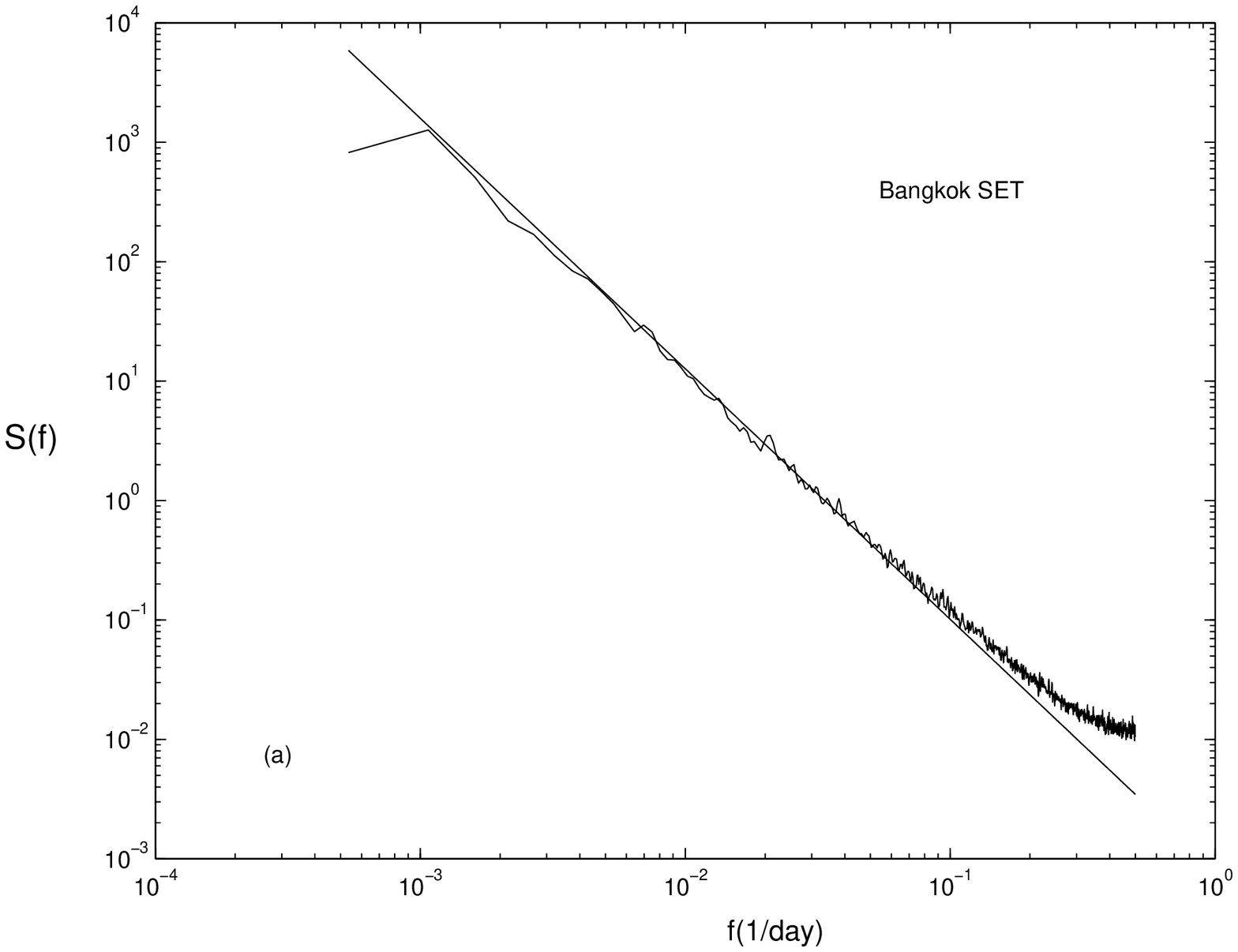,width=3.in,angle=0}}
\mbox{\epsfig{file=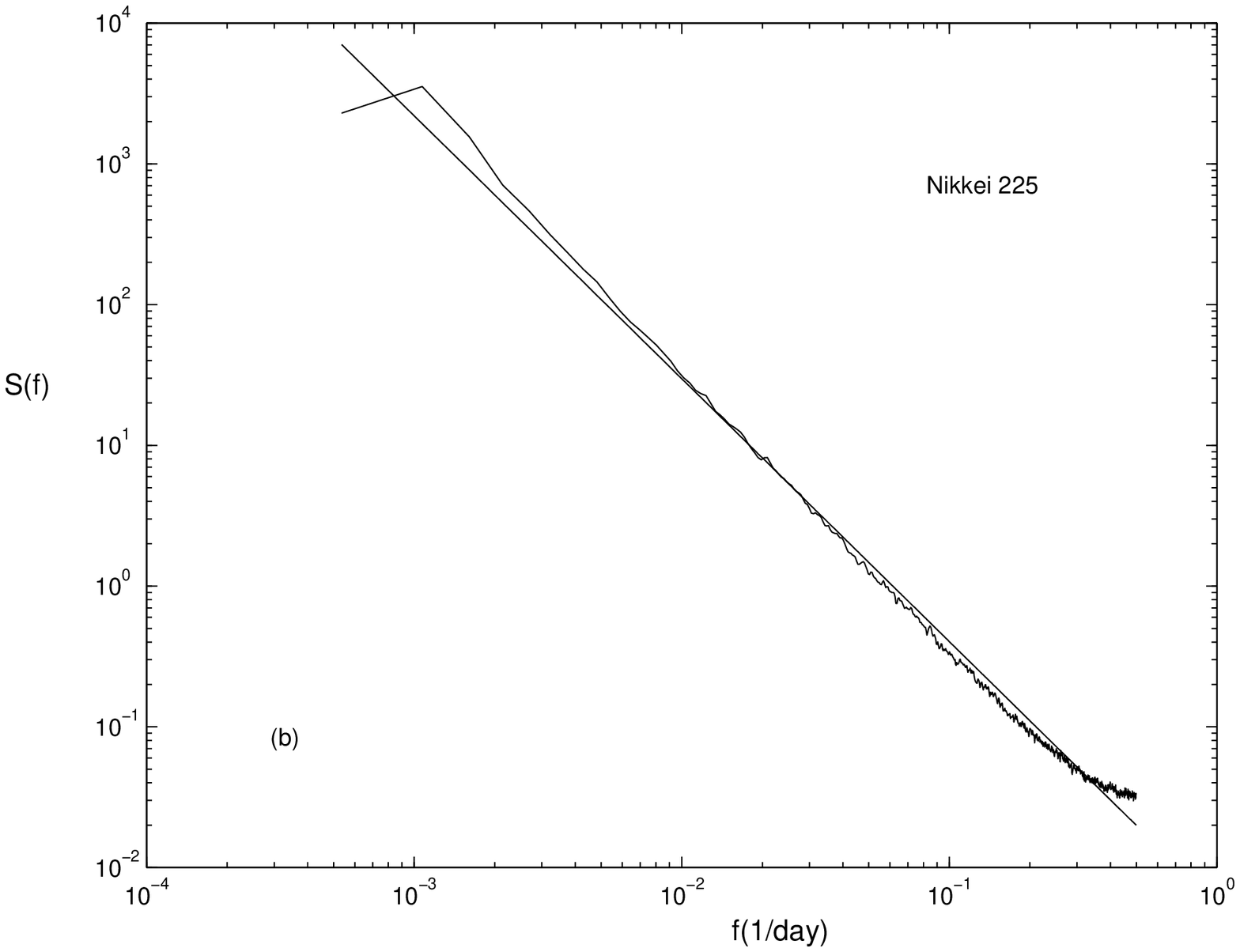,width=3.in,angle=0}}
\end{center}
\caption{The power spectra of the Stock Market indices compared with the behaviour of $f^{-2H(2)-1}$ (straight lines in log-log scale) computed using the Hurst exponents values in the time period $1997$-$2001$; (a) Thailand (Bangkok SET) and (b) JAPAN (Nikkei 225).
The line is the prediction from the generalized Hurst exponent $H(2)$ (Eq. ~\ref{www}).}
\label{f.SpettroSM}
\end{figure}

\begin{figure}
\begin{center}
%\mbox{\epsfig{file=FigSpettroTRER.eps,width=6.in,angle=0}}
\mbox{\epsfig{file=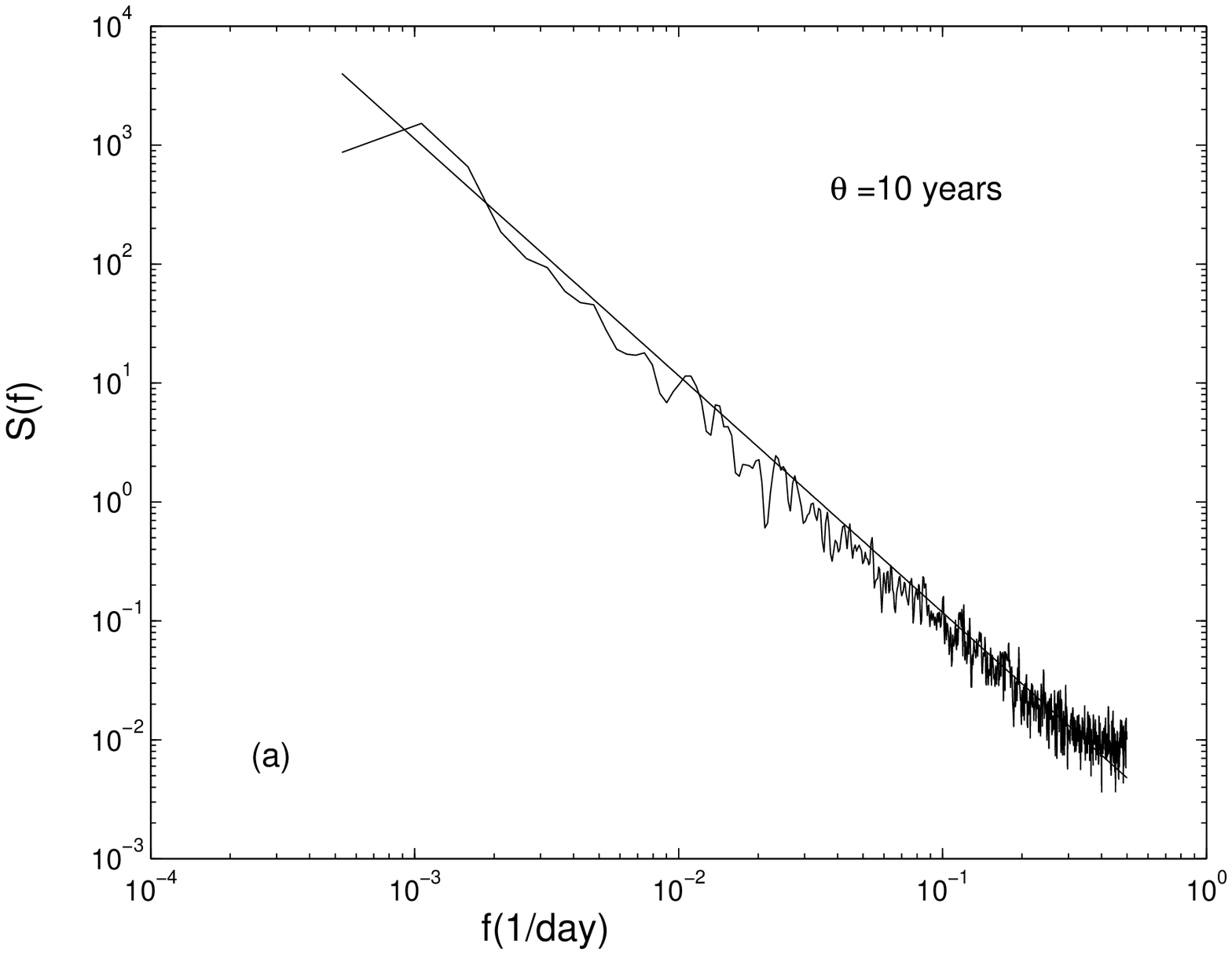,width=3.in,angle=0}}
\mbox{\epsfig{file=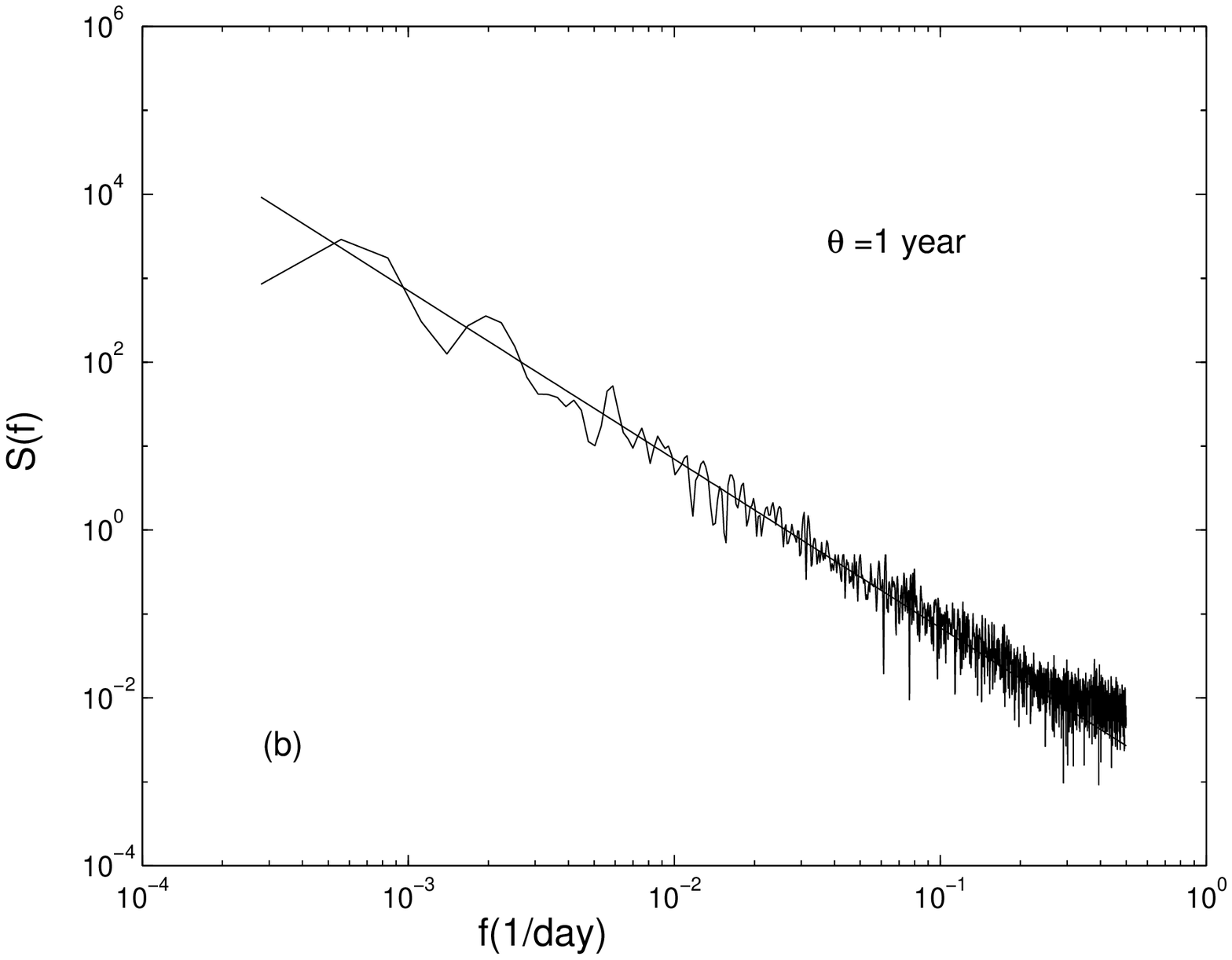,width=3.in,angle=0}}
\end{center}
\caption{The power spectra compared with the behaviour
of $f^{-2H(2)-1}$ (straight lines in log-log scale) computed using the Hurst exponents values in the time period $1997$-$2001$; (a) Treasury rates having maturity dates $\theta=10$ years; (b) Eurodollar rates having maturity dates $\theta=1$ year in the time period $1990$-$1996$.
The line is the prediction from the generalized Hurst exponent $H(2)$ (Eq. ~\ref{www}).}
\label{f.SpettroTRER}
\end{figure}

In order to empirically investigate the statistical properties of
the time series in the frequency domain we perform a spectral
analysis computing the power spectral density (PSD) \cite{Kay} by
using the periodogram approach, that is currently one of the most
popular and computationally efficient PSD estimator. This is a
sensitive way to estimate the limits of the scaling regime of the
data increments. The results for some SM, FX, TR and ER data in
the time periods $1997$-$2001$ and $1990$-$1996$, are shown in
Figs.~\ref{f.SpettroFX},~\ref{f.SpettroSM},~\ref{f.SpettroTRER}.
For SM and FX we compute the power spectra of the logarithm of
these time series. As one can see the power spectra show clear
power law behaviors: $S(f) \sim f^{-\beta }$. This behavior holds
for all the other data.

\begin{figure}
\begin{center}
\mbox{\epsfig{file=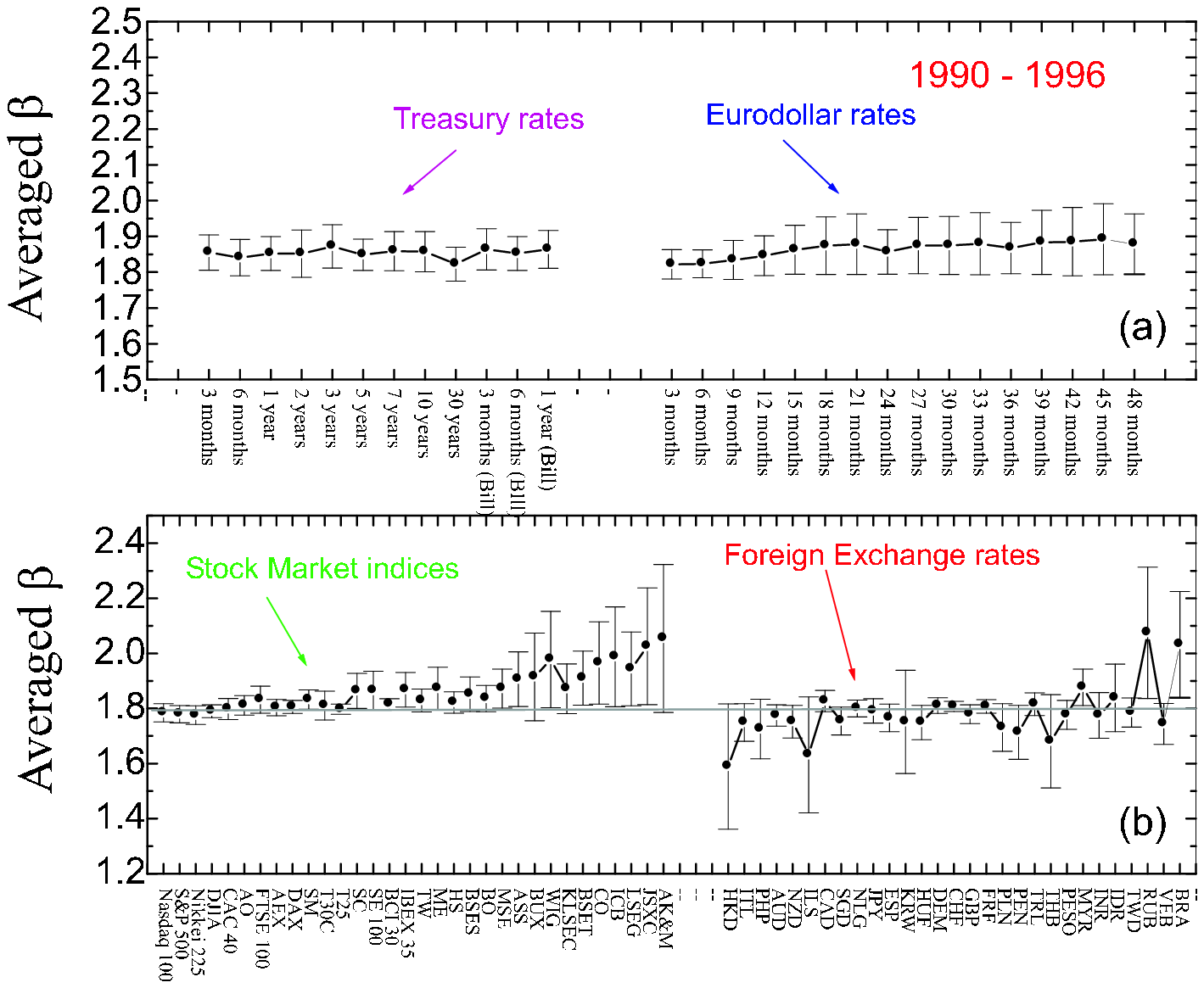,width=6.in,angle=0}}
\end{center}
\caption{(a) The averaged $\beta$ values computed from the power spectra (mean square regression) of the Treasury and Eurodollar rates time series in the period from $1990$ to $1996$; (On the $x$-axis the corresponding maturities dates are reported.) (b) The averaged $\beta$ values computed from the power spectra of the Stock Market indices and Foreign Exchange rates in the time period reported in Tabs. ~\ref{t.1} and ~\ref{t.2}. The horizontal gray line corresponds to the value of $\beta$ obtained from the simulated random walks reported in Table ~\ref{t.HbetaGaussian}. (On the $x$-axis the corresponding data-sets are reported.)}
\label{f.beta}
\end{figure}

The non-stationary features have been investigated by varying the
window-size on which the spectrum is calculated from $100$ days to
up to the entire size of the time series. The power spectra
coefficients are calculated through a mean square regression in
log-log scale. The values reported in Fig.~\ref{f.beta} are the
average of the evaluated $\beta$ over different windows and the
error bars are their standard deviations. Fig.~\ref{f.beta}~(a)
refers to a time period between $1990$ to $1996$ whereas the Stock
Market indices and Foreign Exchange rates (Fig.~\ref{f.beta}~(b))
are analyzed over the time periods reported in
Tables~\ref{t.1}~and~\ref{t.2}. Moreover, the averaged $\beta$
values in a different time period, namely from $1997$ to $2001$
are reported in Table~\ref{t.betatimeperiod9701} for Foreign
Exchange rates, Stock Market indices and Treasury rates. These
values differ from the spectral density exponent expected for a
pure Brownian motion ($\beta=2$) \cite{Feller}. However, we have
shown in Section~\ref{s.metodology} that this method is biased and
we have indeed found power spectra exponents around $1.8$ for
random walks using three different random numbers generators.

It must be noted that, the power spectrum is only a second order
statistic and its slope is not enough to validate a particular
scaling model: it gives only partial information about the
statistics of the process.

\subsection{Comparison between the generalized Hurst exponent and the power spectra}
\label{s.comparison}

We here compare the behavior of the power spectra $S(f)$ with the
function $f^{2H(2)-1}$ which - according to Eq.~\ref{www} - is the
scaling behavior expected in the frequencies domain for a time
series which scales in time with a generalized Hurst exponent
$H(2)$. We performed such a comparison for all the financial data
and we report in Figs.~\ref{f.SpettroFX},~\ref{f.SpettroSM} those
for Foreign Exchange rates and Stock Market indices for Thailand
and JAPAN (in the time period $1997$-$2001$). In
Fig.~\ref{f.SpettroTRER} are reported the comparison for the
Treasury and the Eurodollar rates having maturity dates
$\theta=10$ years and $\theta=1$ year respectively. As one can see
the agreement between the power spectra behavior and the
prediction from the generalized Hurst analysis is very
satisfactory. This result holds also for all the other data. Note
that the values of $2 H(2)+1$ do not in general coincide with the
values for the power spectral exponents evaluated by means of the
mean square regression. The method through the generalized Hurst
exponent appears to be more powerful in catching the scaling
behaviour even in the frequency domain.

\section{Conclusion}
\label{s.conclusion}
Scaling behaviors are rather universal across
financial markets. By analyzing the scaling properties of the
$q$-order moments (Eq.~\ref{multi}) we show that the generalized
Hurst exponent $H(q)$ (Eq.~\ref{Hurstgen}) is a powerful
instrument to characterize and differentiate the structure of such
a scaling properties. We show that $qH(q)$ has a non-linear
dependence on $q$ which is a clear signature of deviations from
pure Brownian motion and other additive models. The empirical
analysis across a wide variety of stock markets shows that the
exponent $H(2)$ is sensitive to the degree of development of the
market. On one end we find: Nasdaq $100$ (US), $S\&P 500$ (US),
Nikkei $225$ (Japan), Dow Jones Industrial Average (US), CAC $40$
(France) and All Ordinaries (Australia); all with $H(2)<0.5$.
Whereas, on the opposite side, we find the Russian AK\&M, the
Indonesian JSXC, the Peruvian LSEG, etc. (Fig. ~\ref{f.H2log}
(b)); all with $H(2)>0.5$. We observe emerging structures also in
the scaling behaviors of interest rates and exchange rates. The
robustness of the present empirical approach is tested in several
ways: by varying the maximum time-step ($\tau_{max}$); by using
the Jackknife method; by varying the time-window sizes; by
comparing with three distinct simulated Brownian motions. We
verify that the observed differentiation among different degrees
of market development is clearly emerging well above the numerical
fluctuations. Finally, from the comparison between the empirical
power spectra and the prediction from the scaling analysis
(Eq.~\ref{www},
Figs.~\ref{f.SpettroFX},~\ref{f.SpettroSM}~and~\ref{f.SpettroTRER})
we show that the method through the generalized Hurst exponent
describes well the scaling behavior even in the frequency domain.

\section*{Acknowledgments}
T. Di Matteo wishes to thank Sandro Pace for fruitful discussions
and support. M. Dacorogna benefited from discussions with the
participants to the CeNDEF workshop in Leiden, June 2002.

\section*{Appendix:The Hurst exponent}

Let us consider a time series $X(t)$ defined at discrete time
intervals $t = \nu,~2\nu,~ 3\nu,~...~k\nu$. Let us define the
average over a period $T$ (which must be an entire multiple of
$\nu$) as
\begin{equation}\label{average}
\left< X \right>_T = \frac{\nu}{T} \sum_{k=1}^{T/\nu} X(k \nu) \;\;\;.
\end{equation}

The difference between the maximum and the minimum values of
$X(t)$ in the interval $[\nu,T]$ is called the range $R$, which is
defined as:
\begin{equation}
\label{Rang}
R(T) =
max [X(t)]_{\nu \le t \le T}
- min [X(t)]_{\nu \le t \le T}
\;\;\;.
\end{equation}
The Hurst exponent $H$ is defined from the scaling property of the ratio:
\begin{equation}
\label{hurst}
\frac{R(T)}{S(T)} \propto \left( \frac{T}{\nu} \right)^{H} \;\;\;,
\end{equation}
where $S(T)$ is the standard deviation:
\begin{equation}
\label{Stdev}
S(T) = \sqrt{\frac{\nu}{T}\sum_{k=1}^{T/\nu}
[X(k \nu)- \left< X \right>_T]^2 }  \;\;\;.
\end{equation}

The Hurst exponent is sensitive to the long-range statistical
dependence in the signal. It was proved by Hurst
\cite{HurstBlack65} and Feller \cite{Feller} that the asymptotic
behavior for any \emph{independent random process} (Poisson
process) with finite variance is given by:
\begin{equation}\label{rr}
\frac{R(T)}{S(T)} = \left(\frac{\pi}{2 \nu}T \right)^{1/2}
\end{equation}
which implies $H = 1/2$. However, many processes in nature are not
independent random processes, but on the contrary show significant
long-term correlations. In this case the asymptotic scaling law is
modified and $R/S$ is asymptotically given by the power law
behavior in Eq.~\ref{hurst} with $H \not= 0.5$. It must be noted
that the lack of robustness of the original Hurst $R/S$ approach
in the presence of short memory, heterocedasticity, multiple scale
behaviors has been largely discussed in the literature (see for
instance \cite{Lo91,Teverovsky99,Przybylowicz2000}) and therefore
several alternative approaches have been proposed. Also the fact
that the range relies on maxima and minima makes the method error
prone because any outlier present in the data would have a strong
influence on the range.

%\newpage
%\listoffigures
%\newpage
%\listoftables

\newpage
\pagestyle{empty}
\vspace*{3cm}

%\newpage

\vspace*{3cm}

\newpage

\vspace*{2cm}

\newpage

\vspace*{2cm}

\newpage

\vspace*{2cm}

\newpage

\vspace*{2cm}
\newpage

\vspace*{1cm}

\newpage

\vspace*{1cm}

\newpage

\vspace*{3cm}

\newpage

\vspace*{3cm}

\newpage
\begin{table}
\caption{Foreign Exchange rates (FX/USD).}
\label{t.1}
\begin{tabular}{cccccc}
$ Country $ & $FX$ & Time period & $ Country $ & $FX$ & Time period \\
\hline
Hong Kong & HKD & 1990-2001 & United Kingdom & GBP & 1990-2001 \\
Italy & ITL & 1993-2001 & France & FRF & 1993-2001 \\
Philippines & PHP & 1991-2001 & Poland & PLN & 1993-2001 \\
Australia & AUD & 1990-2001 & Peru & PEN & 1993-2001 \\
New Zealand & NZD & 1990-2001 & Turkey & TRL & 1992-2001 \\
Israel & ILS & 1990-2001 & Thailand & THB & 1990-2001 \\
Canada & CAD & 1993-2001 & Mexico & PESO & 1993-2001 \\
Singapore & SGD & 1990-2001 & Malaysia & MYR & 1990-2001 \\
Netherlands & NLG & 1993-2001 & India & INR & 1990-2001 \\
Japan & JPY & 1990-2001 & Indonesia & IDR & 1991-2001 \\
Spain & ESP & 1990-2001 & Taiwan & TWD & 1990-2001 \\
South Korea & KRW & 1990-2001 & Russia & RUB & 1993-2001 \\
Hungary & HUF & 1993-2001 & Venezuela & VEB & 1993-2001 \\
Germany & DEM & 1990-2001 & Brazil & BRA & 1993-2001 \\
Switzerland & CHF & 1993-2001 & \\
\end{tabular}
\end{table}

%\newpage
\begin{table}
\caption{Stock Market indices (SM).}
\label{t.2}
\begin{tabular}{ccc}
$ Country $ & $SM$ & Time period \\
\hline
United States & Nasdaq 100 & 1990-2001 \\
United States & S\&P 500 & 1987-2001 \\
Japan & Nikkei 225 & 1990-2001 \\
United States & Dow Jones Industrial Average (DJIA) & 1990-2001 \\
France & CAC 40 & 1993-2001 \\
Australia & All Ordinaries (AO) & 1992-2001 \\
United Kingdom & FTSE 100 & 1990-2001 \\
Netherlands & AEX & 1993-2001 \\
Germany & DAX & 1990-2001 \\
Switzerland & Swiss Market (SM) & 1993-2001 \\
New Zealand & Top 30 Capital (T30C) & 1992-2001 \\
Israel & Telaviv 25 (T25) & 1992-2001 \\
South Korea & Seoul Composite (SC) & 1990-2001 \\
Canada & Toronto SE 100 (SE 100) & 1993-2001 \\
Italy & BCI 30 & 1993-2001 \\
Spain & IBEX 35 & 1990-2001 \\
Taiwan & Taiwan Weighted (TW) & 1990-2001 \\
\end{tabular}
\end{table}

\begin{table}

%\caption{Table 2}
\begin{tabular}{ccc}
Table $2$ (continued)\\
$ Country $ & $SM$ & Time period \\
\hline
Argentina & Merval (ME) & 1993-2001 \\
Hong Kong & Hang Seng (HS) & 1990-2001 \\
India & Bombay SE Sensex (BSES) & 1990-2001 \\
Brazil & Bovespa (BO) & 1993-2001 \\
Mexico & Mexico SE (MSE) & 1993-2001 \\
Singapore & All Singapore Shared (ASS) & 1990-2001 \\
Hungary & Budapest BUX (BUX) & 1993-2001 \\
Poland & Wig (WIG) & 1991-2001 \\
Malaysia & KLSE Composite (KLSEC) & 1990-2001 \\
Thailand & Bangkok SET (BSET) & 1990-2001 \\
Philippines & Composite (CO) & 1990-2001 \\
Venezuela & Indice de Cap. Bursatil (ICB) & 1993-2001 \\
Peru & Lima SE General (LSEG) & 1993-2001 \\
Indonesia & JSX Composite (JSXC) & 1990-2001 \\
Russia & AK\&M Composite (AK\&M) & 1993-2001 \\
\end{tabular}
\end{table}

\newpage
\begin{table}
\caption{Treasury rates ($TR_i({\theta})$).}
\label{t.3}
\begin{tabular}{cccccccccccc}
$ i $ & $\theta$ & $ i $ & $\theta$ \\
\hline
1 & 3 months & 7 & 7 years  \\
2 & 6 months & 8 & 10 years \\
3 & 1 year & 9 & 30 years   \\
4 & 2 years & 10 & 3 months (Bill)   \\
5 & 3 years & 11 & 6 months (Bill) \\
6 & 5 years & 12 & 1 year (Bill) \\
\end{tabular}
\end{table}

\newpage
\begin{table}
\caption{Eurodollar rates ($ER_i({\theta})$).}
\label{t.4}
\begin{tabular}{cccccccccccc}
$ i $ & $\theta$ & $ i $ & $\theta$ \\
\hline
1 & 3 months & 9 & 27 months \\
2 & 6 months & 10 & 30 months   \\
3 & 9 months & 11 & 33 months \\
4 & 12 months & 12 & 36 months \\
5 & 15 months & 13 & 39 months \\
6 & 18 months & 14 & 42 months \\
7 & 21 months & 15 & 45 months \\
8 & 24 months & 16 & 48 months \\
\end{tabular}
\end{table}

\newpage

\begin{table}
\caption{Hurst exponents $H(1)$ and $H(2)$ and averaged $\beta$ values computed for random walks simulated by using three different random numbers generators: 1) Randn=Normally distributed random numbers with mean $0$ and variance $1$; 2) Rand=Uniformly distributed random numbers in the interval ($0,1$) and 3) Normrnd=Random numbers from the normal distribution with mean $0$ and standard deviation $1$. These are average values on $100$ simulations of random walks with $991$ and $3118$ numbers of data points.}
\label{t.HbetaGaussian}
\begin{tabular}{cccccccccccc}
$ N $ & $ H(1) $ & $ H(2) $ & $\beta$ \\
\hline
1) Randn \\
$991$ & $ 0.50 \pm 0.01 $ & $0.50 \pm 0.01 $ & $ 1.8 \pm 0.1 $ \\
$3118$ & $ 0.50 \pm 0.01 $ & $0.50 \pm 0.01 $ & $ 1.80 \pm 0.03 $ \\
2) Rand \\
$991$ & $ 0.50 \pm 0.01 $ & $ 0.49 \pm 0.01 $ & $ 1.8 \pm 0.1 $ \\
$3118$ & $ 0.47 \pm 0.01 $ & $ 0.50 \pm 0.01 $ & $ 1.80 \pm 0.03 $ \\
3) Normrnd \\
$991$ & $ 0.49 \pm 0.01 $ & $0.49 \pm 0.01 $ & $ 1.8 \pm 0.1 $ \\
$3118$ & $ 0.50 \pm 0.01 $ & $0.50 \pm 0.01 $ & $ 1.80 \pm 0.03 $ \\
\end{tabular}
\end{table}

\newpage

\begin{table}
\caption{Hurst exponents $H(1)$ and $H(2)$ for Foreign Exchange rates, Stock Market indices and Treasury rates in the time period from $1997$ to $2001$.}
\label{t.Htimeperiod9701}
\begin{tabular}{cccccccccccc}
$ Data $ & $ H(1) $ & $ H(2)$ & $ Data $ & $ H(1) $ & $ H(2)$ \\
\hline
Foreign Exchange rates \\
HKD & $0.41 \pm 0.01$ & $0.34 \pm 0.01$ & GBP & $0.50 \pm 0.02$ & $0.48 \pm 0.02$    \\
ITL & $0.51 \pm 0.01$ & $0.51 \pm 0.01$  & FRF & $0.51 \pm 0.01$ & $0.51 \pm 0.01$   \\
PHP & $0.52 \pm 0.01$ & $0.43 \pm 0.02$  & PLN & $0.54 \pm 0.01$ & $0.50 \pm 0.01$  \\
AUD & $0.52 \pm 0.01$ & $0.502 \pm 0.002$ & PEN & $0.52 \pm 0.01$ & $0.41 \pm 0.03$ \\
NZD & $0.49 \pm 0.01$ & $0.48 \pm 0.01$ &  TRL & $0.56 \pm 0.01$ & $0.44 \pm 0.04$    \\
ILS & $0.48 \pm 0.02$ & $0.47 \pm 0.02$ & THB & $0.53 \pm 0.01$ & $0.50 \pm 0.02$   \\
CAD & $0.51 \pm 0.01$ & $0.48 \pm 0.01$  & PESO & $0.53 \pm 0.01$ & $0.50 \pm 0.01$  \\
SGD & $0.50 \pm 0.01$ & $0.47 \pm 0.03$ & MYR & $0.51 \pm 0.03$ & $0.45 \pm 0.05$    \\
NLG & $0.51 \pm 0.01$ & $0.51 \pm 0.01$ &  INR & $0.58 \pm 0.02$ & $0.53 \pm 0.01$   \\
JPY & $0.50 \pm 0.01$ & $0.49 \pm 0.01$ &  IDR & $0.56 \pm 0.03$ & $0.53 \pm 0.03$  \\
ESP & $0.50 \pm 0.01$ & $0.49 \pm 0.01$ & TWD & $0.58 \pm 0.01$ & $0.51 \pm 0.01$  \\
KRW & $0.50 \pm 0.03$ & $0.39 \pm 0.06$ & RUB & $0.64 \pm 0.02$ & $0.47 \pm 0.03$  \\
HUF & $0.52 \pm 0.01$ & $0.52 \pm 0.01$   & VEB & $0.54 \pm 0.04$ & $0.49 \pm 0.02$    \\
DEM & $0.51 \pm 0.01$ & $0.51 \pm 0.01$  & BRA & $0.59 \pm 0.02$ & $0.60 \pm 0.01$  \\
CHF & $0.51 \pm 0.01$ & $0.50 \pm 0.01$   \\
\end{tabular}
\end{table}

\begin{table}
\begin{tabular}{cccccccccccc}
Table $6$ (continued)\\
$ Data $ & $ H(1) $ & $ H(2)$ & $ Data $ & $ H(1) $ & $ H(2)$ \\
\hline
Stock Market indices\\
Nasdaq 100 & $0.47\pm 0.01$ & $0.45\pm 0.01$  & TW & $0.53 \pm 0.01$ & $0.51 \pm 0.01$   \\
S\&P 500 & $0.47 \pm 0.02$ & $0.44 \pm 0.01$  & ME & $0.57 \pm 0.01$ & $0.53 \pm 0.01$  \\
Nikkei 225 & $0.46 \pm 0.01$ & $0.43 \pm 0.01$ & HS & $0.53 \pm 0.01$ & $0.49 \pm 0.01$   \\
DJIA & $0.49 \pm 0.01$ & $0.464 \pm 0.004$ & BSES & $0.54 \pm 0.01$ & $0.52 \pm 0.01$  \\
CAC 40 & $0.47 \pm 0.02$ & $0.46 \pm 0.02$  & BO & $0.51 \pm 0.01$ & $0.48 \pm 0.01$  \\
AO & $0.49 \pm 0.02$ & $0.46 \pm 0.03$ & MSE & $0.57 \pm 0.01$ & $0.52 \pm 0.01$ \\
FTSE 100 & $0.46 \pm 0.02$ & $0.44 \pm 0.01$ & ASS & $0.57 \pm 0.01$ & $0.54 \pm 0.02$ \\
AEX & $0.49 \pm 0.01$ & $0.47 \pm 0.02$ & BUX & $0.52 \pm 0.01$ & $0.49 \pm 0.01$ \\
DAX & $0.50 \pm 0.01$ & $0.47 \pm 0.01$ & WIG & $0.49 \pm 0.01$ & $0.44 \pm 0.01$  \\
SM & $0.50 \pm 0.02$ & $0.48 \pm 0.02$& KLSEC & $0.60 \pm 0.01$ & $0.51 \pm 0.02$  \\
T30C & $0.49 \pm 0.01$ & $0.46 \pm 0.01$  & BSET & $0.59 \pm 0.01$ & $0.55 \pm 0.01$\\
T25 & $0.53 \pm 0.01$ & $0.51 \pm 0.01$& CO & $0.59 \pm 0.01$ & $0.54\pm 0.01$    \\
SC & $0.53 \pm 0.01$ & $0.51 \pm 0.01$  &   ICB & $0.61 \pm 0.02$ & $0.55 \pm 0.02$      \\
SE 100 & $0.51 \pm 0.01$ & $0.48 \pm 0.01$& LSEG & $0.61 \pm 0.01$ & $0.58 \pm 0.01$  \\
BCI 30 & $0.52 \pm 0.01$ & $0.48 \pm 0.01$& JSXC & $0.57 \pm 0.02$ & $0.53 \pm 0.02$      \\
IBEX 35 & $0.50 \pm 0.01$ & $0.48 \pm 0.01$& AK\&M & $0.65 \pm 0.03$ & $0.51 \pm 0.01$     \\
\end{tabular}
\end{table}

\begin{table}
\begin{tabular}{cccccccccccc}
Table $6$ (continued)\\
$ Data $ & $ H(1) $ & $ H(2)$ \\
\hline
Treasury rates\\
$TR_1$ & $0.48 \pm 0.01$ & $0.44 \pm 0.02$     \\
$TR_2$ & $0.55 \pm 0.01$ & $0.52\pm 0.02$  \\
$TR_3$ & $0.54 \pm 0.01$ & $0.52 \pm 0.02$    \\
$TR_4$ & $0.53 \pm 0.01$ & $0.52 \pm 0.02$    \\
$TR_5$ & $0.52 \pm 0.01$ & $0.50 \pm 0.01$    \\
$TR_6$ & $0.51 \pm 0.02$ & $0.49 \pm 0.01$   \\
$TR_7$ & $0.49 \pm 0.02$ & $0.48 \pm 0.01$     \\
$TR_8$ & $0.52 \pm 0.01$ & $0.50 \pm 0.02$     \\
$TR_9$ & $0.51 \pm 0.01$ & $0.48 \pm 0.01$    \\
$TR_{10}$ & $0.51 \pm 0.01$ & $0.48 \pm 0.02$       \\
$TR_{11}$ & $0.56 \pm 0.01$ & $0.54 \pm 0.02$   \\
$TR_{12}$ & $0.55 \pm 0.01$ & $0.53 \pm 0.02$   \\
\end{tabular}
\end{table}
\newpage

\begin{table}
\caption{The averaged $\beta$ values computed from the power spectra of the Stock Market indices, Foreign Exchange rates and Treasury rates in the time period from $1997$ to $2001$.}
\label{t.betatimeperiod9701}
\begin{tabular}{cccccccccccc}
$ Data $ & Averaged $\beta$ & $ Data $ & Averaged $\beta$\\
\hline
Foreign Exchange rates \\
HKD & $1.6 \pm 0.2$ & GBP & $1.79 \pm 0.03$    \\
ITL & $1.80 \pm 0.03$ & FRF & $1.81 \pm 0.04$ \\
PHP & $1.8 \pm 0.1$ & PLN & $1.79 \pm 0.04$    \\
AUD & $1.8 \pm 0.1$ & PEN & $1.6 \pm 0.2$    \\
NZD & $1.8 \pm 0.1$ & TRL & $1.7 \pm 0.1$    \\
ILS & $1.8 \pm 0.1$ & THB & $1.83 \pm 0.03$   \\
CAD & $1.80 \pm 0.03$ & PESO & $1.81 \pm 0.04$    \\
SGD & $1.81 \pm 0.02$ & MYR & $1.8 \pm 0.1$    \\
NLG & $1.81 \pm 0.04$ & INR & $1.8 \pm 0.1$   \\
JPY & $1.9 \pm 0.1$ & IDR & $1.83 \pm 0.04$    \\
ESP & $1.80 \pm 0.04$ & TWD & $1.8 \pm 0.1$ \\
KRW & $1.8 \pm 0.1$ & RUB & $2.1 \pm 0.3$    \\
HUF & $1.80 \pm 0.03$ & VEB & $1.8 \pm 0.1$   \\
DEM & $1.81 \pm 0.03$ & BRA & $2.0 \pm 0.2$   \\
CHF & $1.8 \pm 0.1$ \\
\end{tabular}
\end{table}

\begin{table}
\begin{tabular}{cccccccccccc}
Table $7$ (continued)\\
$ Data $ & Averaged $\beta$ & $ Data $ & Averaged $\beta$\\
\hline
Stock Market indices\\
Nasdaq 100 & $1.7 \pm 0.1$ & TW & $1.9 \pm 0.1$    \\
S\&P 500 & $1.8 \pm 0.1$ & ME & $1.8 \pm 0.1$    \\
Nikkei 225 & $1.8 \pm 0.1$ & HS & $1.8 \pm 0.1$    \\
DJIA & $1.80 \pm 0.03$ & BSES & $1.82 \pm 0.03$    \\
CAC 40 & $1.8 \pm 0.1$ & BO & $1.80 \pm 0.02$    \\
AO & $1.8 \pm 0.1$ & MSE & $1.9 \pm 0.1$    \\
FTSE 100 & $1.81 \pm 0.03$ & ASS & $1.9 \pm 0.1$    \\
AEX & $1.8 \pm 0.1$ & BUX & $1.82 \pm 0.04$    \\
DAX & $1.8 \pm 0.1$ & WIG & $1.8 \pm 0.1$    \\
SM & $1.8 \pm 0.1$ & KLSEC & $1.8 \pm 0.1$    \\
T30C & $1.8 \pm 0.1$ & BSET & $1.9 \pm 0.1$    \\
T25 & $1.9 \pm 0.1$ & CO & $2.0 \pm 0.2$    \\
SC & $1.9 \pm 0.1$ & ICB & $2.0 \pm 0.2$    \\
SE 100 & $1.9 \pm 0.1$ & LSEG & $2.0 \pm 0.2$    \\
BCI 30 & $1.9 \pm 0.1$ & JSXC & $1.9 \pm 0.1$    \\
IBEX 35 & $1.8 \pm 0.1$ & AK\&M & $1.9 \pm 0.2$    \\
%\end{tabular}
%\end{table}
%\begin{table}
%\begin{tabular}{cccccccccccc}
%Table $7$ (continued)\\
%$ Data $ & Averaged $\beta$ & $ Data $ & Averaged $\beta$\\
%\hline
Treasury rates\\
$TR_1$ & $1.8 \pm 0.1$ & $TR_7$ & $1.9 \pm 0.1$    \\
$TR_2$ & $1.83 \pm 0.04$ & $TR_8$ & $1.9 \pm 0.1$    \\
$TR_3$ & $1.86 \pm 0.05$ & $TR_9$ & $1.8 \pm 0.1$    \\
$TR_4$ & $1.88 \pm 0.06$ & $TR_{10}$ & $1.82 \pm 0.04$    \\
$TR_5$ & $1.9 \pm 0.1$ & $TR_{11}$ & $1.85 \pm 0.04$    \\
$TR_6$ & $1.9 \pm 0.1$ & $TR_{12}$ & $1.9 \pm 0.1$    \\
\end{tabular}
\end{table}


\newpage
\begin{thebibliography}{00}

\bibitem{Muller90}
M\"{u}ller, U. A., M. M. Dacorogna, R. B. Olsen, O. V. Pictet, M.
Schwarz and C. Morgenegg, 1990, Statistical study of foreign
exchange rates, empirical evidence of a price change scaling law,
and intraday analysis,
 Journal of Banking and Finance 14, 1189-1208.

\bibitem{BouchaudPot}
Bouchaud, J. P. and M. Potters, 1997, Th\'eorie des Risques
Financiers (Alea, Saclay).


\bibitem{Rebonato}
Rebonato, R., 1998, Interest-rate option models (John Wiley \& Sons, New
York).

\bibitem{Wilmott}
Wilmott, P.,  S. Howison, J. Dewynne, 1999, The mathematics of financial
Derivatives (Cambridge University Press, Cambridge).


\bibitem{Hull}
Hull, J., 2000, Options, futures, and other derivatives (Prentice
Hall, New York).

\bibitem{LibrMant}
Mantegna, R. N. and H. E. Stanley, 2000, An Introduction to
Econophysics (Cambridge University Press, Cambridge).

\bibitem{LibDac}
Dacorogna, M. M., R. Gen\c{c}ay, U. A. Muller, R. Olsen, O. V.
Pictet, 2001, An Introduction to High-Frequency Finance (Academic
Press).

\bibitem{Mandelbrot}
Mandelbrot, B. B., 1963, The variation of Certain Speculative Prices,
Journal of Business 36, 394-419.

\bibitem{MandelbrotB}
Mandelbrot, B. B., 1997, Fractals and Scaling in Finance (Springer-Verlag,
New York).

\bibitem{Groenendijk}
Groenendijk, P. A., A. Lucas and C. G. de Vries, 1998, A Hybrid Joint Moment Ratio Test for Financial Time Series, Preprint of the Erasmus University, obtainable at http://www.few.eur.nl/few/people/cdevries/.

\bibitem{Corsi01}
Corsi, F., G. Zumbach, U. A. M\"{u}ller, M. M. Dacorogna, 2001,
Consistent High-Precision Volatility from High-Frequency Data,
 Economic Notes, Review of Banking, Finance and Monetary Economics 30 (2), 183-204.

\bibitem{MantSta95}
Mantegna, R. and H. E. Stanley, 1995, Scaling behavior in the dynamics of an economic index, Nature 376, 46-49.

\bibitem{Ballocchi}
Ballocchi, G., M. M. Dacorogna, R. Gen\c{c}ay and B. Piccinato, 1999, Intraday Statistical Properties of Eurofutures,
 Derivatives Quarterly 6 (2), 28-44.

\bibitem{Daco01}
Dacorogna, M. M., U. A. M\"{u}ller, R. B. Olsen and O. V. Pictet,
2001, Defining efficiency in heterogeneous markets, Quantitative
Finance 1 (2), 198-201.

\bibitem{Bachelier1900}
Bachelier, L., 1900, Theory of Speculation (Translation of 1900 French edn), in (edr), P. H. Cootner , 1964, The Random Character of Stock Market Prices, (The MIT Press, Cambridge, MA) 17-78.

\bibitem{Osborne59}
Osborne, M. F., 1959, Brownian motion in the stock market, Operations Research 7, 145-173.

\bibitem{Mand65}
Mandelbrot, B. B., 1965, Une classe de processus stochastiques homoth\'etiques \`a soi; application \`a la loi climatologique de H. E. Hurst, Comptes Rendus (Paris) 260, 3274-3277.

\bibitem{Mand68}
Mandelbrot, B. B. and Van Ness, 1968, Fractional Brownian motions, fractional noises and applications, SIAM Review 10, 422-437.

\bibitem{Mand62}
Mandelbrot, B. B., 1962, Sur certains prix sp\'eculatifs:faits empiriques et mod\`ele bas\'e sur des processus stables additifs de Paul L\'evy, Comptes Rendus (Paris) 254, 3968-3970.

\bibitem{Fama63}
Fama, E. F., 1963, Mandelbrot and the stable Paretian hypothesis, Journal of Business 36, 420-429.

\bibitem{Fama65}
Fama, E. F., 1965, The behavior of stock-market prices,
 Journal of Business 38, 34-105.

\bibitem{Mand67}
Mandelbrot, B. B., 1967, The variation of some other speculative prices, Journal of Business 40, 393-413.

\bibitem{Mirowski95}
Mirowski, P., 1995, Mandelbrot's economics after a quarter century, Fractals 3, 581-600.

\bibitem{Evertsz95}
Evertsz, C., 1995, Fractal geometry of financial time series, Fractals 3, 609-616.

\bibitem{BlakeLeBaron01}
LeBaron, B., 2001, Stochastic volatility as a simple generator of apparent financial power laws and long memory, Quantitative Finance 1, 621-631.

\bibitem{Lux01}
Lux, T., 2001, Turbulence in financial markets: the surprising explanatory power of simple cascade models, Quantitative Finance 1, 632-640.

\bibitem{Mandelbrot01}
Mandelbrot, B. B., 2001, Scaling in financial prices: IV. Multi-fractal concentration, Quantitative Finance 1 (6), 641-649.

\bibitem{StanPlerou}
Stanley, H. E. and V. Plerou, 2001, Scaling and universality in economics: empirical results and theoretical interpretation, Quantitative Finance 1 (6), 563-567.

\bibitem{Bouchaud2000}
Bouchaud, J.-P., M. Potters and M. Meyer, 2000, Apparent multi-fractality in financial time series, European Physical Journal B 13, 595-599.

\bibitem{Stanle96}
Stanley, H. E. et al., 1996, Can statistical physics contribute to the science of economics?, Fractals 4, 415-425.

\bibitem{ScalEco}
Brock, W. A., 1999, Scaling in Economics: a Reader's Guide, Industrial and Corporate Change 8 (3), 409-446.

\bibitem{DiMatteo}
Di Matteo, T. and T. Aste, 2002, How does the Eurodollars interest rate behave?,
 International Journal of Theoretical and Applied Finance 5, 107-122.

\bibitem{HurstBlack65}
Hurst, H. E., 1951, Long-term storage capacity of reservoirs, Transaction of the American Society of Civil Engineers 116, 770-808; Hurst, H. E., R. Black, Y. M. Sinaika, 1965, Long-Term Storage in Reservoirs: An experimental Study (Constable, London).

\bibitem{Lo91}
Lo, A., 1991, Long-Term memory in stock market prices, Econometrica 59, 1279-1313.

\bibitem{Peng94}
Peng, C.-K., S. V. Buldyrev, S. Havlin, M. Simons, H. E. Stanley, A. L. Goldberger, 1994, Mosaic organization of DNA nucleodites, Physical Review E 49, 1685-1689.

\bibitem{Stanley96}
Stanley, H. E., S. V. Buldyrev, A. L. Goldberger, S. Havlin, C.-K. Peng, M. Simmons, 1999, Scaling Features of Noncoding DNA, Physica A 273, 1-18.

\bibitem{Viswa97}
Viswanathan, G. M., S. V. Buldyrev, S. Havlin, H. E. Stanley, 1997, Quantification of DNA patchiness using long-range correlation measures, Biophysical Journal 72, 866-875.

\bibitem{Hu01}
Hu, K., P. Ch. Ivanov, Z. Chen, P. Carpena and H. E. Stanley, 2001, Effect of trends on detrended fluctuation analysis, Physical Review E 64, 011114.

\bibitem{Vandewalle97}
Vandewalle, N. and M. Ausloos, 1997, Coherent and Random Sequences in Financial Fluctuations, Physica A 246, 454-459.

\bibitem{Ausloos2000}
Ausloos, M., 2000, Statistical physics in foreign exchange currency and stock markets, Physica A 285, 48-65.

\bibitem{Geweke83}
Geweke, J. and S. Porter-Hudak, 1983, The estimation and application of long-memory time series models, Journal of Time Series Analysis 4, 221-238.

\bibitem{Sowell92}
Sowell, F. B., 1992, Maximum likelihood estimation of stationary univariate fractionally integrated time series models, Journal of Econometrics 53, 165-188.

\bibitem{Carles2000}
Grau-Carles, P., 2000, Empirical evidence of long-range correlations in stock returns, Physica A 287, 396-404.

\bibitem{Zipf}
Zipf, G. K., 1949, Human Behavior and the Principle of Least Effort (Addisson-Wesley, Cambridge, MA).

\bibitem{Ellinger71}
Ellinger, A. G., 1971, The Art of Investment (Bowers \& Bowers, London).

\bibitem{Mehrabi97}
Mehrabi, A. R., H. Rassamdana, and M. Sahimi, 1997, Characterization of long-range correlations in complex distributions and
 profiles, Physical Review E 56, 712-722.

\bibitem{Simonsen98}
Simonsen, I., A. Hansen, and O. Nes, 1998, Determination of the Hurst exponent by use of wavelet trasforms, Physical Review E 58, 2779-2787.

\bibitem{Percival2000}
Percival, D. B. and A. T. Walden, 2000, Wavelet Methods for Time Series Analysis (Cambridge University Press, Cambridge).

\bibitem{Gencay2001}

Gen\c{c}ay, R., F. Sel\c{c}uk and B. Whitcher, 2001, Scaling
properties of foreign exchange volatility, Physica A 289, 249-266.

\bibitem{Vandewalle98}
Vandewalle, N., M. Ausloos, 1998, Sparseness and roughness of foreign exchange rates, International Journal of Modern Physics C 9, 711-720.


\bibitem{VandAus98}
Vandewalle, N., M. Ausloos, 1998, Crossing of two mobile averages: A method for measuring the roughness exponent, Physical Review E 58, 6832-6834.

\bibitem{Ivanova99}
Ivanova, K., M. Ausloos, 1999, Low q-moment multifractal analysis of Gold price, Dow Jones Industrial Average and BGL-USD,
 European Physical Journal B 8, 665-669.

\bibitem{Barabasi91}
Barabasi, A. L., T. Vicsek, 1991, Multifractality of self-affine fractals, Physical Review A 44, 2730-2733.

\bibitem{Feder88}
Feder, J., 1988, Fractals (Plenum Press, New York).

\bibitem{West85}
West, B. J., 1985, The Lure of Modern Science: Fractal thinking,
(World Scientific).

\bibitem{Flandrin89}
Flandrin, P., 1989, On the spectrum of fractional Brownian motions, IEEE Transaction on Information Theory 35, 197-199.

\bibitem{1AuIva01}
Ausloos, M., K. Ivanova, 2001, False Euro (FEUR) exchange rate correlated behaviors and investment strategy,
 European Physical Journal B 20, 537-541.

\bibitem{2AuIva01}
Ausloos, M., K. Ivanova, 2001, Correlation Between Reconstructed EUR Exchange Rates Versus CHF, DKK, GBP, JPY and USD,
 International Journal of Modern Physics C 12, 169-195.

\bibitem{VanAus98}
Vandewalle, N., M. Ausloos, 1998, Multi-affine analysis of typical currency exchange rates, European Physical Journal B 4, 257-261.

\bibitem{kunsch89}
Kunsch, H. R., 1989, The Jackknife and the Bootstrap for General
Stationary Observations , The Annals of Statistics 17, 1217-1241.

\bibitem{Kay}
Kay, S. M. and S. L. Marple, 1981, Spectrum Analysis-A modern Perspective, Proceedings of the IEEE 69, 1380-1415.

\bibitem{Feller}
Feller, W., 1971, An Introduction to Probability Theory and its Applications (Wiley, New York).

\bibitem{Teverovsky99}
Teverovsky, V.,  M. U. Taquu, W. Willinger, 1999, A critical look at Lo's modified R/S statistic, Journal of Statistical Planning an Inference 80, 211-227.

\bibitem{Przybylowicz2000}
Weron, R. and B. Przybylowsciz, 2000, Hurst analysis of electricity price dynamics, Physica A 283, 462-468.


\end{thebibliography}
\end{document}